\documentclass[prd,aps,twocolumn,a4paper,showpacs,floatfix,nofootinbib]{revtex4-1}

\usepackage{mathrsfs}
\usepackage{amsmath,amsfonts,amssymb}
\usepackage{latexsym}
\usepackage{bm}
\usepackage[dvips]{graphicx}
\usepackage{color}
\usepackage{multirow}

\def\de{\partial}
\def\lm{{\ell m}}

\def\ii{{\rm i}}

\def\r{{\hat{r}}}

\def\F{{\cal F}}

\def\O{{\cal O}}

\newcommand{\be}{\begin{equation}}  
\newcommand{\ee}{\end{equation}}
\newcommand{\bea}{\begin{eqnarray}}           
\newcommand{\eea}{\end{eqnarray}} 
\newcommand{\beqn}{\begin{eqnarray*}}
\newcommand{\eeqn}{\end{eqnarray*}}
\newcommand{\ba}{\begin{align}}
\newcommand{\ea}{\end{align}}

\newcommand{\fig}[2]{\caption{\label{#1} #2}}

\begin{document}

\title{Binary black hole merger in the extreme-mass-ratio limit: a multipolar analysis.}

\author{Sebastiano \surname{Bernuzzi}$^1$}
\author{Alessandro \surname{Nagar}$^2$}
\affiliation{$^1$Theoretical Physics Institute, University of Jena, 07743 Jena, Germany}
\affiliation{$^2$Institut des Hautes Etudes Scientifiques, 91440 Bures-sur-Yvette, France}

\date{\today}

\begin{abstract}
  Building up on previous work, we present a new calculation of the 
  gravitational wave emission generated during the 
  transition from quasi-circular inspiral to plunge, merger and ringdown
  by a binary system of nonspinning black holes, of masses 
  $m_1$ and $m_2$, in the extreme mass ratio limit, $m_1 m_2\ll (m_1+m_2)^2$.
  The relative dynamics of the system is computed 
  {\it without making any adiabatic approximation} by using an effective 
  one body (EOB) description, namely by representing the
  binary by an effective particle of 
  mass $\mu=m_1 m_2/(m_1+m_2)$ moving in a \hbox{(quasi-)Schwarzschild}
  background of mass $M=m_1+m_2$ and submitted to an $\O(\nu)$ 
  5PN-resummed analytical radiation reaction force, with $\nu=\mu/M$.
  The gravitational wave emission is calculated via a multipolar 
  Regge-Wheeler-Zerilli type perturbative approach (valid in
  the limit $\nu\ll 1$).
  We consider three mass ratios, $\nu=\{10^{-2},10^{-3},10^{-4}\}$,
  and we compute the multipolar waveform up to $\ell=8$.
  We estimate energy and angular momentum losses during the
  quasiuniversal and quasigeodesic part of the plunge phase
  and we analyze the structure of the ringdown.
  We calculate the gravitational recoil, or ``kick'', imparted 
  to the merger remnant by the gravitational wave emission and
  we emphasize the importance of higher multipoles to get a final 
  value of the recoil $v/(c\nu^2)=0.0446$.
  We finally show that there is an {\it excellent fractional agreement} 
  ($\sim 10^{-3}$) (even during the plunge) between the 5PN EOB 
  analytically-resummed radiation reaction flux and the numerically computed 
  gravitational wave angular momentum flux. This is a further confirmation 
  of the aptitude of the EOB formalism to accurately model 
  extreme-mass-ratio inspirals, as needed for the future space-based
  LISA gravitational wave detector.
\end{abstract}

\pacs{
   04.30.Db,  
    04.25.Nx,  
    95.30.Sf,  
 }
 
\maketitle

\section{Introduction}
%
%
After the breakthroughs of
2005~\cite{Pretorius:2005gq,Campanelli:2005dd,Baker:2005vv}, 
state-of-the-art numerical relativity (NR) codes can nowadays 
routinely evolve (spinning) coalescing binary black hole
systems with comparable masses and extract the gravitational wave (GW)
signal with high accuracy~\cite{Hannam:2007ik,Hannam:2007wf,
Sperhake:2008ga,Campanelli:2008nk,Scheel:2008rj,
Chu:2009md,Mosta:2009rr,Pollney:2009yz}. 
However, despite these considerable improvements, 
the numerical computation of coalescing black hole binaries where 
the mass ratio is considerably different from 1:1 is 
still challenging. To date, the mass ratio 10:1 (without spin) 
remains the highest that was possible to numerically evolve
through the transition from inspiral to plunge and merger with
reasonable accuracy~\cite{Gonzalez:2008bi,Lousto:2010tb}. 
In recent years, work at the interface between analytical 
and numerical relativity, notably using the effective-one-body (EOB)
resummed analytical formalism~\cite{Buonanno:1998gg,Buonanno:2000ef,
Damour:2000we,Damour:2001tu,Buonanno:2005xu,Damour:2009ic},
has demonstrated the possibility of using NR results 
to develop accurate analytical models of dynamics and 
waveforms of coalescing black-hole binaries~\cite{Buonanno:2007pf,
Damour:2007yf,Damour:2007vq,Damour:2008te,Boyle:2008ge,
Buonanno:2009qa,Damour:2009kr,Pan:2009wj,Barausse:2009xi}.

%
%
By contrast, when the mass ratio is large, approximation methods
based on black-hole perturbation theory are expected to yield 
accurate results, therefore enlarging our black-hole binaries
knowledge by a complementary perspective. 
In addition, when the larger black-hole mass is in the range 
$10^5M_\odot$-$10^7M_\odot$, the GWs emitted by the radiative 
inspiral of the small object fall within the sensitivity band 
of the proposed space-based detector LISA~\cite{lisa1,lisa2}, 
so that an accurate modelization of these extreme-mass-ratio-inspirals 
(EMRI) is one of the goals of current GW research.
   
The first calculation of the complete gravitational waveform emitted 
during the transition from inspiral to merger in the 
extreme-mass-ratio limit was performed in
Refs.~\cite{Nagar:2006xv,Damour:2007xr}, thanks to the combination of 
2.5PN Pad\'e resummed radiation reaction force~\cite{Damour:1997ub} 
with Regge-Wheeler-Zerilli perturbation
theory~\cite{Regge:1957td,Zerilli:1970se,Nagar:2005ea,Martel:2005ir}.
This test-mass laboratory was then used to understand, 
element by element, the physics that enters in the dynamics 
and waveforms during the transition from inspiral to plunge 
(followed by merger and ringdown), providing important inputs 
for EOB-based analytical models. In particular, it helped to:
(i) discriminate between two expressions of the resummed radiation 
reaction force; 
(ii) quantify the accuracy of the resummed multipolar 
wavefom; 
(iii) quantify the effect of non-quasi-circular corrections
(both to waveform and radiation reaction); 
(iv) qualitatively understand 
the process of generation of quasi-normal modes (QNMs); and 
(v) improve the matching procedure of the ``insplunge'' waveform to a
``ringdown'' waveform with (several) QNMs. 
In the same spirit, the multipolar expansion of the 
gravitational wave luminosity of a test-particle in
circular orbits on a Schwarzschild 
background~\cite{Cutler:1993vq,Yunes:2008tw,Pani:2010em} 
was helpful to devise an improved resummation 
procedure~\cite{Damour:2007yf,Damour:2008gu} 
of the PN (Taylor-expanded) multipolar 
waveform~\cite{Kidder:2007rt,Blanchet:2008je}. Such resummation 
procedure is one of the cardinal elements of what we think is 
presently the best EOB analytical 
model~\cite{Damour:2009kr,Pan:2009wj}.
Similarly, Ref.~\cite{Yunes:2009ef} compared  ``calibrated'' 
EOB-resummed waveforms~\cite{Boyle:2008ge} with Teukolsky-based 
perturbative waveforms, and confirmed that the EOB framework 
is well suited to model EMRIs for LISA.

In addition, recent numerical achievement in the calculation of the 
conservative gravitational self-force (GSF) of circular orbits 
in a Schwarzschild background~\cite{Detweiler:2008ft,Barack:2009ey,Barack:2010tm} 
prompted the interplay between post-Newtonian (PN) and 
GSF efforts~\cite{Blanchet:2009sd,Blanchet:2010zd}, and 
EOB and GSF efforts~\cite{Damour:2009sm}.
In particular, the information coming from GSF data helped 
to break the degeneracy (among some EOB parameters) which was left
after using comparable-mass NR data to constrain the EOB
formalism~\cite{Damour:2009sm}. (See also Ref.~\cite{Barausse:2009xi}
for a different way to incorporate GSF results in EOB).

%
%
In this paper we present a revisited computation of the
GWs emission from the transition from inspiral to plunge in the 
test-mass limit. We improve the previous calculation of 
Nagar et al.~\cite{Nagar:2006xv} in two aspects: 
one numerical and the other analytical. The first is that 
we use a more accurate (4th-order) numerical 
algorithm to solve the Regge-Wheeler-Zerilli equations numerically;
this allows us to capture the higher order multipolar information 
(up to $\ell=8$) more accurately than in~\cite{Nagar:2006xv}.
The second aspect is that we have replaced the 
2.5PN Pad\'e resummed radiation reaction force 
of~\cite{Nagar:2006xv} with the 5PN resummed one that relies 
on the results of Ref.~\cite{Damour:2008gu}. 

The aim of this paper is then two-fold. 
On the one hand, our new test-mass perturbative allows us to
describe in full, and with high accuracy, the properties of
the gravitational radiation emitted during the transition 
inspiral-plunge-merger and ringdown, {\it without making the adiabatic 
approximation} which is the hallmark of most existing approaches to
the GW emission by EMRI systems~\cite{Glampedakis:2002cb,Hughes:2005qb}. 
We compute the multipolar waveform up to $\ell=8$, we 
discuss the relative weight
of each multipole during the nonadiabatic plunge phase, 
and describe the structure of the ringdown.
In addition, from the multipolar waveform we compute also 
the total recoil, or kick, imparted to the system by the 
wave emission, thereby complementing NR results~\cite{Gonzalez:2008bi}.
On the other hand, we can use our upgraded test-mass laboratory
to provide inputs for the EOB formalism, notably for completing 
the EOB multipolar waveform during the late-inspiral, plunge and
merger. As a first step in this direction, we show that the
analytically resummed radiation reaction 
introduced in~\cite{Damour:2008gu,Damour:2009kr}
gives an {\it excellent fractional agreement} ($\sim 10^{-3}$) 
with the angular momentum flux computed \'a la Regge-Wheeler-Zerilli
even during the plunge phase. 

This paper is organized as follows. In Sec.~II we give a summary 
of the formalism employed. In Sec.~III we describe the multipolar
structure of the waveforms; details  on the energy and angular 
momentum emitted during the plunge-merge-ringdown transition 
are presented as well as an analysis of the ringdown phase. 
Section~IV is devoted to present the computation of the final kick, 
emphasizing the importance of high multipoles. 
The following Sec.~V is devoted to some consistency checks: on the 
one hand, we discuss the aforementioned  agreement between the 
mechanical angular momentum loss and GW energy flux during the 
plunge; on the other hand, we investigate the influence
of EOB ``self-force'' terms (either in the conservative and non 
conservative part of the dynamics)
on the waveforms. We present a summary of our findings in Sec.~VI.
In Appendix~A we supply some technical details related to our 
numerical framework, while in Appendix~B we list some useful numbers.
We use geometric units with $c=G=1$.

\section{Analytic framework}
\label{sec:dynamics}

\subsection{Relative dynamics}

The relative dynamics of the system is modeled specifying the 
EOB dynamics to the small-mass limit. The formalism that we use
here is the specialization to the test-mass limit of the improved
EOB formalism introduced in Ref.~\cite{Damour:2009kr} that crucially 
relies on the ``improved resummation'' procedure of the multipolar 
waveform of Ref.~\cite{Damour:2008gu}.
Let us recall that the EOB approach to the general relativistic 
two-body dynamics is a {\it nonperturbatively resummed}  analytic 
technique which has been developed in  
Refs.~\cite{Buonanno:1998gg,Buonanno:2000ef,Damour:2000we,Damour:2001tu,Buonanno:2005xu}.
This technique uses, as basic input,  the results of PN theory,
such as: (i) PN-expanded equations of motion for two pointlike bodies,
 (ii) PN-expanded radiative multipole moments, and (iii) PN-expanded
energy and angular momentum fluxes at infinity. For the moment, the
most accurate such results are the 3PN conservative 
dynamics~\cite{Damour:2001bu,Blanchet:2003gy}, the 3.5PN energy 
flux~\cite{Blanchet:2001aw,Blanchet:2004bb,Blanchet:2004ek} for the $\nu\neq0$
case, and 5.5PN~\cite{Tanaka:1997dj} accuracy for the $\nu=0$ case.
Then the EOB approach ``packages'' this PN-expanded information in special 
{\it resummed} forms which extend the validity of the PN results 
beyond the expected weak-field-slow-velocity regime into (part of) 
the strong-field-fast-motion regime.
In the EOB approach the relative dynamics of a binary 
system of masses $m_1$ and $m_2$ is described
by a Hamiltonian $H_{\rm EOB}(M,\mu)$ and a radiation reaction force 
${\cal F}_{\rm EOB}(M,\mu)$, where $M\equiv m_1+m_2$ and $\mu\equiv m_1m_2/M$. 
In the general comparable-mass case $H_{\rm EOB}$
has the structure $H_{\rm EOB}(M,\mu)=M\sqrt{1+2\nu(\hat{H}_\nu - 1)}$
where $\nu\equiv \mu/M\equiv m_1m_2/(m_1+m_2)^2$ is the symmetric mass ratio.
In the test mass limit that we are considering, $\nu\ll 1$, we can expand
$H_{\rm EOB}$ in powers of $\nu$. After subtracting inessential constants
we get a Hamiltonian per unit ($\mu$) mass 
$\hat{H}=\lim_{\nu \to 0}(H-{\rm const.})/\mu=\lim_{\nu\to 0}\hat{H}_\nu$.
As in Refs.~\cite{Nagar:2006xv,Damour:2007xr}, we replace the Schwarzschild 
radial coordinated $r_*=r + 2M\log[r/(2M)-1]$ and, correspondingly, the radial
momentum $P_R$ by the conjugate momentum $P_{R_*}$ of $R_*$, so that
the specific Hamiltonian has the form
\begin{equation}
\hat{H} = \sqrt{ A\left( 1+ \frac{p_{\varphi}^2}{\hat{r}^2} \right)+p_{r_*}^2} \ .
\end{equation}
Here we have introduced dimensionless variables $\r\equiv R/M$,$\r_*\equiv
R_*/M$,  $p_{r_*}\equiv P_{R_*}/\mu$, $p_\varphi \equiv P_\varphi/(\mu M)$
and $A=1-2/\r$. Hamilton's canonical equations for $(\r,r_*,p_{r_*},p_{\varphi})$ 
in the equatorial plane ($\theta=\pi/2$) yield
\begin{align}
\label{eob:1}
\dot{\r}_*       &= \dfrac{p_{\r_*}}{\hat{H}} \ , \\
\label{eob:2}
\dot{\r}         &= \dfrac{A}{\hat{H}}p_{r_*}                \equiv v_r \ , \\
\label{eob:3}
\dot{\varphi}   &= \dfrac{A}{\hat{H}}\dfrac{p_\varphi}{\r^2} \equiv \Omega \ ,\\
\label{eob:4}
\dot{p}_{r_*}   &= -\dfrac{\r-2}{\r^3\hat{H}}\left[p_{\varphi}^2\left(\dfrac{3}{\r^2}-\dfrac{1}{\r}\right)+1\right] \ , \\
\label{eob:5}
\dot{p}_\varphi &= \hat{\cal F}_\varphi \ .
\end{align}
Note that the quantity $\Omega$ is dimensionless and represents the orbital frequency
in units of $1/M$.
In these equations the extra term $\hat{\cal F}_{\varphi}$ 
[of order $O(\nu)$] represents the non conservative part of 
the dynamics, namely the radiation reaction force.
Following~\cite{Damour:2006tr,Nagar:2006xv,Damour:2007xr}, we use the
following expression:
\be
\label{eq:rr}
\hat{\F}_\varphi \equiv -\dfrac{32}{5}\nu \Omega^5 \r^4 \hat{f}_{\rm DIN}(v_\varphi),
\ee
where  $v_\varphi = \hat{r}\Omega$ is the azimuthal velocity and 
$\hat{f}_{\rm DIN}=F^{\ell_{\rm max}}/F_{\rm Newt}$ denotes the (Newton normalized) 
energy flux up to multipolar order $\ell_{\rm max}$(in the $\nu=0$ limit) 
resummed according to the ``improved resummation'' technique 
of Ref.~\cite{Damour:2008gu}. This resummation procedure is based on a
particular multiplicative decomposition of the multipolar gravitational
waveform. The energy flux is written as
\be
F^{\ell_{\rm max}}=\dfrac{1}{8\pi}\sum_{\ell =2}^{\ell_{\rm max}} \sum_{m=1}^{\ell} (m\Omega)^2|r h_{\ell m}|^2.
\ee
where $h_{\ell m}$ is the factorized waveform of~\cite{Damour:2008gu}, 
\be
h_{\ell m}=h_{\ell m}^{(N,\epsilon)}\hat{S}_{\rm eff}^{(\epsilon)}T_{\ell m}e^{\ii \delta_{\ell m}}\rho^\ell_{\lm}
\ee
where $h_{\ell m}^{(N,\epsilon)}$ represents the Newtonian contribution
given by Eq.~(4) of~\cite{Damour:2008gu}, $\epsilon=0$ (or $1$) for 
$\ell+m$ even (odd), $\hat{S}^{\epsilon}_{\rm eff}$ is the effective
``source'', Eqs.~(15-16) of~\cite{Damour:2008gu}; $T_{\ell m}$ is
the ``tail factor'' that resums an infinite number of ``leading logarithms''
due to tail effects, Eq.~(19) of~\cite{Damour:2008gu}; $\delta_{\ell m}$ 
is a residual phase correction, Eqs.~(20-28) of~\cite{Damour:2008gu};
and $\rho_{\ell m}$ is the residual modulus 
correction, Eqs.~(C1-C35) in~\cite{Damour:2008gu}.
In our setup we truncate the sum on $\ell$ at $\ell_{\rm max}=8$.
We refer the reader to Fig.~1 (b) of~\cite{Damour:2008gu} 
to figure out the capability of the new resummation
procedure to reproduce the actual flux (computed numerically) for
the sequence of circular orbits in Schwarzshild\footnote{We mention that, although 
Ref.~\cite{Damour:2008gu} also proposes to further (Pad\'e) resum 
the residual amplitude corrections $\rho_{\ell m}$ 
(and, in particular, the dominant  one, $\rho_{22}$) to 
improve the agreement with the ``exact'' data, we prefer 
not to include any of these sophistications here. 
This is motivated by the fact that, 
along the sequence of circular orbits, all the different 
choices are practically equivalent up to (and sometimes below) 
the adiabatic last stable orbit (LSO) at $r=6M$ (see in this respect their Fig.~5). 
In practice our $\rho_{22}$ actually corresponds to the Taylor-expanded 
version (at 5PN order) of the remnant amplitude correction,
denoted $T_5[\rho_{22}]$ in~\cite{Damour:2008gu}.}.

\subsection{Gravitational wave generation}
\label{sec:rwz}

The computation of the gravitational waves
generated by the relative dynamics follows th
same line of Refs.~\cite{Nagar:2006xv,Damour:2007xr}, 
and relies on the numerical solution, in the time
domain, of the Regge-Wheeler-Zerilli equations for
metric perturbations of the Schwarzschild black hole
with a point-particle source. 
Once the dynamics from Eqs.~\eqref{eob:1}-\eqref{eob:5} 
is computed, one needs to solve numerically (for each multipole 
$(\ell,m)$ of even (e) or odd (o) type) 
a couple of decoupled partial differential 
equations
\begin{equation}
\label{eq:rwz}
\de_t^2\Psi^{(\rm e/o)}_{\ell m}-\de_{r_*}^2\Psi^{(\rm e/o)}_{\ell m} + V^{(\rm
  e/o)}_{\ell}\Psi^{(\rm e/o)}_{\ell m} = S^{(\rm e/o)}_{\ell m} \ 
\end{equation}
with source terms $S^{(\rm e/o)}_{\ell m}$ linked to the dynamics of 
the binary. Following~\cite{Nagar:2006xv}, the sources are written 
in the functional form
\begin{align}
\label{source:standard}
S^{(\rm e/o)}_{\ell m} & = G^{(\rm e/o)}_{\ell m}(r,t)\delta(r_*-R_*(t)) \nonumber \\
& + F^{(\rm e/o)}_{\ell m}(r,t)\de_{r_*}\delta(r_*-R_*(t)) \ ,
\end{align} 
with $r$-dependent [rather than $R(t)$-dependent] coefficients 
$G(r)$ and $F(r)$. The explicit expression of the sources is 
given in Eqs.~(20-21) of~\cite{Nagar:2006xv}, to which we 
address the reader for further technical details.
We mention, however, that in our approach the distributional 
$\delta$-function is approximated by a narrow 
Gaussian of finite width $\sigma\ll M$. In Ref.~\cite{Nagar:2006xv} 
it was already pointed out that, if $\sigma$ is sufficiently
small and the resolution is sufficiently high (so that the 
Gaussian can be cleanly resolved) this approximation  is 
competitive with other approaches that employ a mathematically  
more rigorous treatment of the 
$\delta$-function~\cite{Lousto:1997wf,Martel:2001yf,Sopuerta:2005gz}
(see in this respect Table~1 and Fig.~2 of Ref.~\cite{Nagar:2006xv}).
That analysis motivates us to use the same representation 
of the $\delta$-function also in this paper, but together 
with an improved numerical algorithm to solve the wave equations.
In fact, the solution of Eqs.~\eqref{eq:rwz}
is now provided via the method of lines by means of a 4th-order Runge-Kutta
algorithm with 4th-order finite differences used to approximate the space
derivatives. This yields better accuracy in the waveforms (using resolutions
comparable to those of Ref.~\cite{Nagar:2006xv}), and allows to better
resolve the higher multipoles. More details about the numerical implementation, 
convergence properties, accuracy, and comparison with published results 
are given in Appendix~\ref{sec:numerical}.

From the numerically calculated master functions $\Psi^{(\rm e/o)}_{\ell m}$, 
one can then obtain, when considering the limit $r\to\infty$, the $h_+$ and $h_\times$ 
gravitational-wave polarization amplitude
\begin{align}
\label{eq:hplus_cross}
h_+-{\rm i}h_{\times} = \dfrac{1}{r}\sum_{\ell\geq 2,m}\sqrt{\frac{(\ell+2)!}{(\ell-2)!}}
     \left(\Psi^{(\rm e)}_{\ell m}
     +{\rm i}\Psi^{(\rm o)}_{\ell m}\right)
     \;_{-2}Y^{\ell m} \ ,
\end{align}
where $\;_{-2}Y^{\ell m}\equiv\,_{-2}Y^{\ell m}(\theta,\varphi)$ are 
the $s=2$ spin-weighted spherical harmonics~\cite{goldberg67}. 
From this expression, all the interesting second-order quantities
follow. The emitted power,
\be
\label{eq:dEdt}
\dot{E}  = \frac{1}{16\pi}\sum_{\ell\geq 2,m}\frac{(\ell+2)!}{(\ell-2)!}
	\left(\left|\dot{\Psi}^{(\rm o)}_{\ell m}\right|^2 + 
        \left|\dot{\Psi}^{(\rm e)}_{\ell m}\right|^2\right)\;, \\
\ee
the angular momentum flux
\be
\label{eq:dJdt}
\dot{J}  =  \frac{1}{32\pi}\sum_{\ell\geq 2,m}\bigg\{{\rm i}m\frac{(\ell+2)!}{(\ell -2)!}
\left[\dot{\Psi}^{(\rm e)}_{\ell m}\Psi^{(\rm e)*}_{\ell m}
+\dot{\Psi}_{\ell m}^{({\rm o})}\Psi^{(\rm o)*}_{\ell m}\right] + c.c.\bigg\} \\
\ee
and the linear momentum flux~\cite{Thorne:1980ru,Pollney:2007ss,Ruiz:2007yx}
\begin{align}
\label{eq:dPdt}
\F^{\bf P}_x + \ii\F^{\bf P}_y &= \dfrac{1}{8\pi}\sum_{\ell\geq 2,m}
\bigg[\ii a_{\lm} \dot{\Psi}_{\lm}^{({\rm e})}\dot{\Psi}^{({\rm o})*}_{\ell,m+1}\nonumber\\
&+b_{\lm}\left(\dot{\Psi}^{(\rm e)}_{\lm}\dot{\Psi}^{(\rm
    e)*}_{\ell+1,m+1}+\dot{\Psi}^{(\rm o)}_{\lm}\dot{\Psi}^{(\rm o)*}_{\ell
    +1,m+1}\right)\bigg] \ .
\end{align}
with
\begin{align}
a_{\lm}    & = 2(\ell-1)(\ell+2)\sqrt{(\ell-m)(\ell+m+1)}\\
b_{\ell m} & =\dfrac{(\ell +3)!}{(\ell +1)(\ell -2)!}\sqrt{\dfrac{(\ell + m
    +1)(\ell+m+2)}{(2\ell+1)(2\ell +3)}} .
\end{align}

\section{Relative dynamics and waveforms}
\label{res:waves}
\begin{figure}[t]
\begin{center}
  \includegraphics[width=0.35\textwidth]{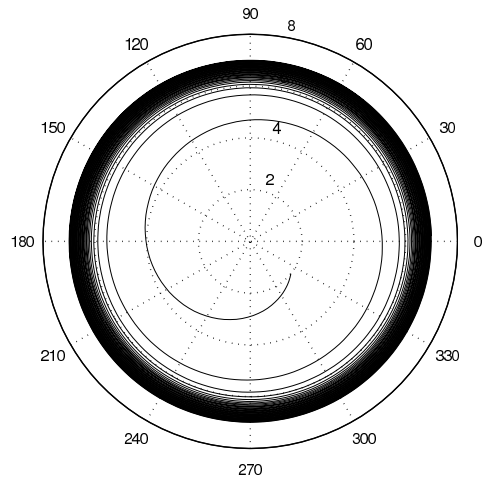}
  \caption{  \label{fig:trj} Transition from quasicircular inspiral 
  orbit to plunge. Initial position is $r_0=7M$ and $\nu=10^{-3}$.}
\end{center} 
\end{figure}

Let us now consider the dynamics and waveforms obtained within our
new setup. Evidently, at the {\it qualitative} level our results are 
analogous to those of Refs.~\cite{Nagar:2006xv,Damour:2007xr}.
By contrast, at the {\it quantitative} level, dynamics and waveforms
are slightly different due to the new, more accurate, radiation reaction force.
The particle is initially at $r=7M$. The dynamics is initiated with the 
so called post-circular  initial data for ($p_\varphi$,$p_{r}$) 
introduced in Ref.~\cite{Buonanno:2000ef} and specialized to the $\mu\to 0$ 
limit (see Eqs.~(9)-(13) of Ref.~\cite{Nagar:2006xv}). Because of the
smallness of the value of $\mu$ we are using, this approximation
is sufficient to guarantee that the initial eccentricity is 
negligible. To have a better modelization of the extreme-mass-ratio
limit regime we considered three values of the mass ratio $\nu$, 
namely $\nu=\{10^{-2},\, 10^{-3},\,10^{-4}\}$.
The values of $\nu$ are chosen so that the particle passes 
through a long (when $\nu\leq 10^{-3}$) quasicircular adiabatic 
inspiral before entering the nonadiabatic plunge phase.
Fig~\ref{fig:trj} displays the relative trajectory for
$\nu=10^{-3}$. The system executes about 40 orbits 
before crossing the LSO at $r=6M$ while plunging 
into the black hole.

\begin{figure*}[t]
\begin{center}
  \includegraphics[width=0.9\textwidth]{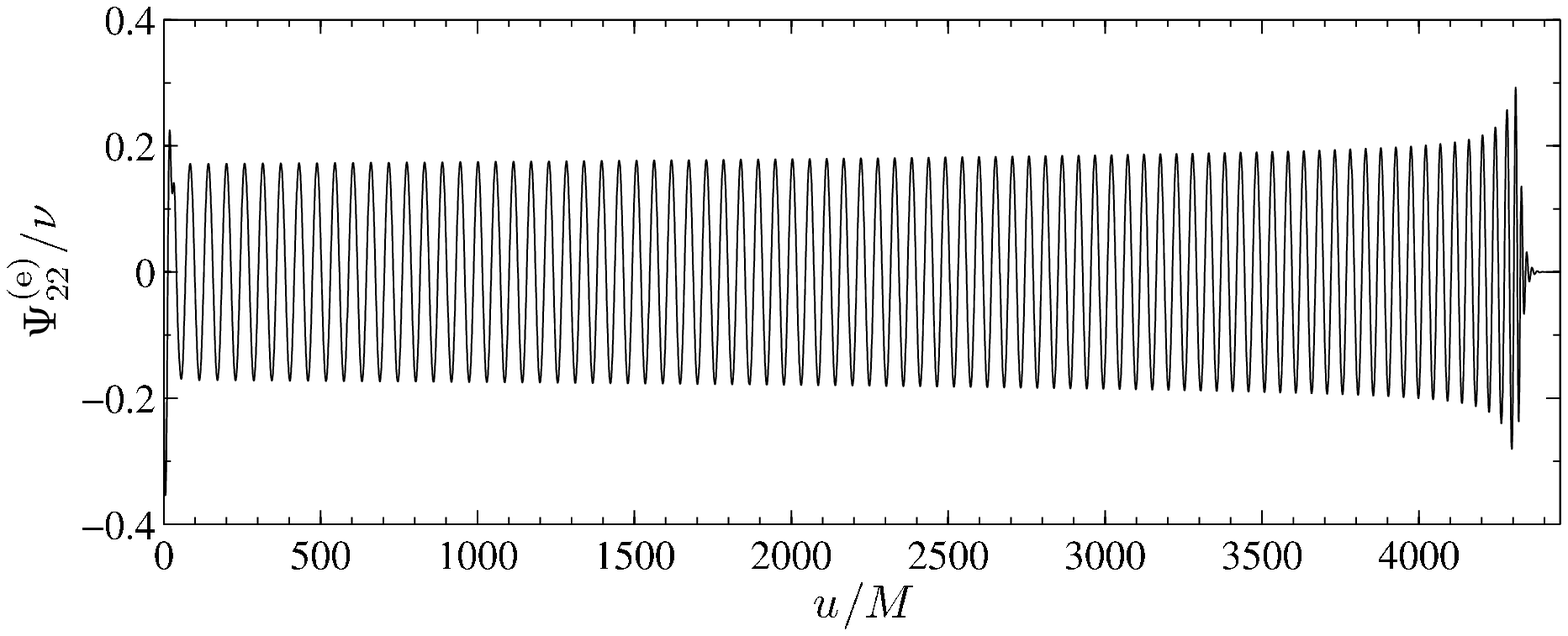}
  \fig{fig:full_wave}{Complete $\ell=m=2$ gravitational (Zerilli)
   waveform corresponding to the dynamics depicted in Fig.~\ref{fig:trj}.
   The waveform is extracted at $r_*^{\rm obs}/M=1000$.}
\end{center} 
\end{figure*}

The main multipolar contribution to the gravitational signal is clearly 
the $\ell=m=2$. The real part of the corresponding  
waveform is displayed in Fig.~\ref{fig:full_wave}. It is 
extracted at $r_*^{\rm obs}=1000M$ and it is shown 
versus observer's retarded time $u=(t^{\rm obs}-r_*^{\rm obs})M$.
Note how the amplitude of the long wavetrain emitted during the 
adiabatic quasicircular inspiral grows  very slowly for about 
$4000M$, until the transition from inspiral to plunge 
around the crossing of the adiabatic LSO frequency. 
In the following we want however to focus on the higher 
order multipolar contributions to the waveform, as they are 
particularly relevant in our test-mass setup. The computation 
of these multipoles and their inclusion in the analysis is 
one of the new results of this paper
\footnote{We note that calculations up to $\ell=4$ were already performed
  in Ref.~\cite{Nagar:2006xv,Damour:2007xr}, but 
  no higher-order multipolar waveforms were either shown or discussed
  in details. The present calculations rely strongly on the new 
  developed 4th-order code. An explicit comparison between the two 
  codes is shown in Appendix~\ref{sec:numerical}.
}.

Figure~\ref{fig:module_multipoles} summarizes the information a
bout the multipolar waveforms up to $\ell=8$.
The left panels show the moduli (normalized by the mass ratio $\nu$),
while the right panels show the corresponding instantaneous gravitational
wave frequencies $M\omega_{\ell m}$.
We show, for each value of $\ell$, the dominant
(even-parity) ones, i.e. those with $m=\ell$, together with some
subdominant (odd-parity) ones. The comparison between the moduli 
highlights how the amplitude of higher modes, that is almost 
negligible during the adiabatic inspiral, can be magnified of about 
factor two (see the $\ell=2$, $m=1$ case) or three 
(see the $m=\ell=6$ case) during the nonadiabatic 
plunge phase. This fact is expected to have some relevance in
those computations that are dominated by the nonadiabatic
plunge phase, like the computation of the recoil velocity
imparted to the center of mass of the system due to the
linear momentum carried away by GWs~\cite{Damour:2006tr}.
As we will see in Sec.~\ref{sec:recoil}, high-order multipoles
are, in fact, needed to obtain an accurate result. 
An analysis of the relative importance of the different multipoles
based on energy considerations is the subject of
Sec.~\ref{sec:qgeoplunge}. 

\begin{figure*}[t]
\begin{center}
  \includegraphics[width=0.45\textwidth]{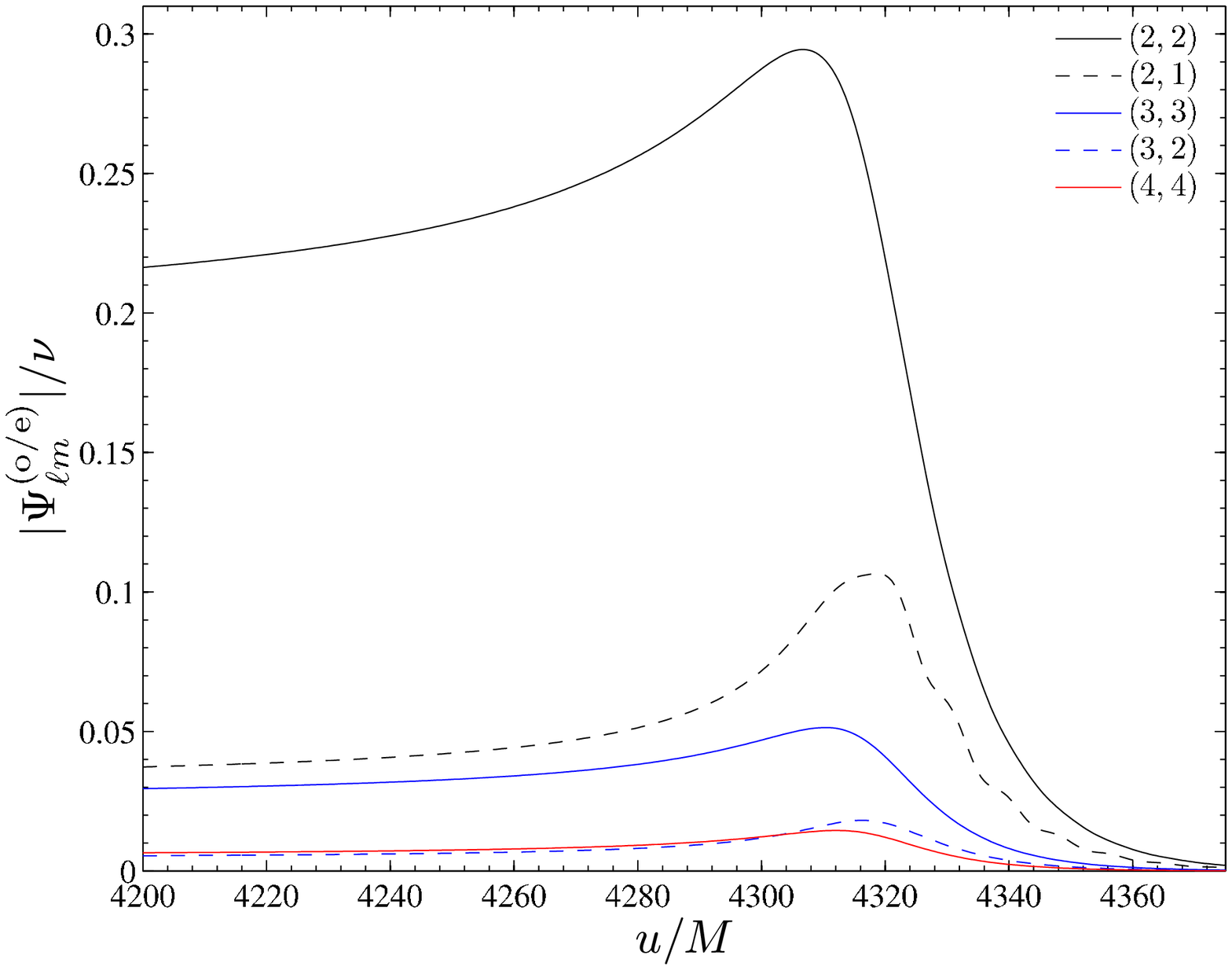}\qquad
  \includegraphics[width=0.45\textwidth]{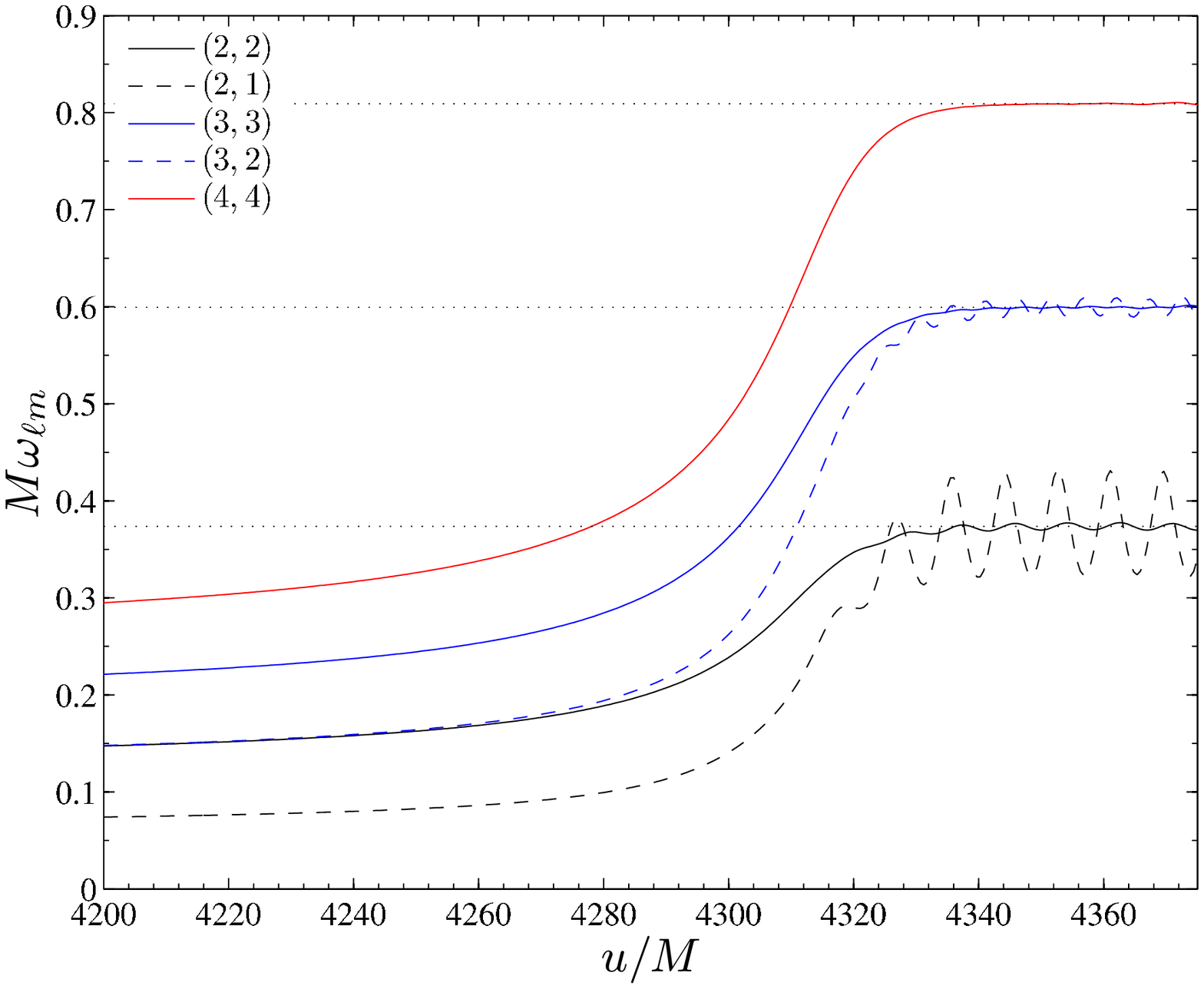}\\
  \vspace{5 mm}
  \includegraphics[width=0.45\textwidth]{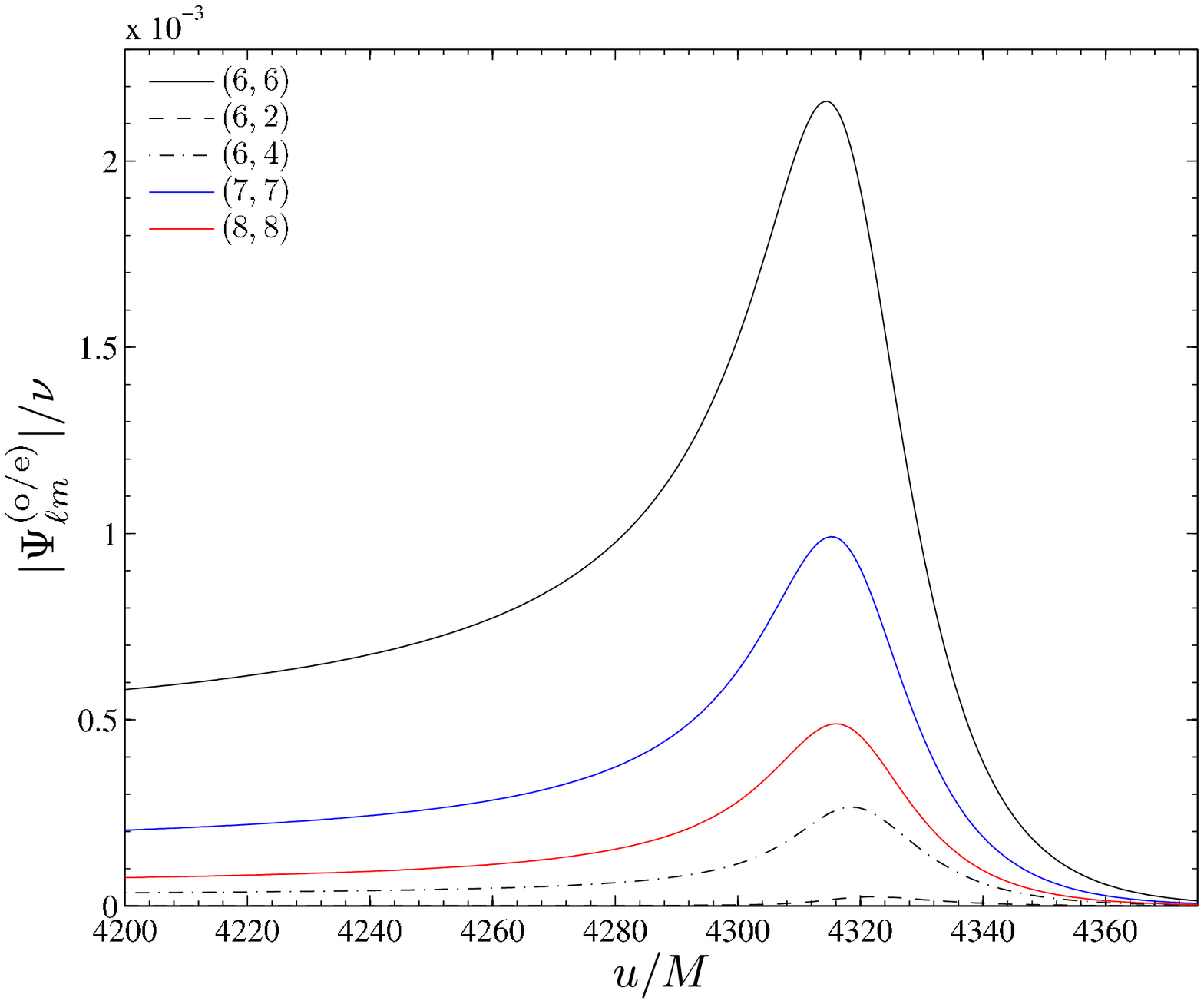}\qquad
  \includegraphics[width=0.45\textwidth]{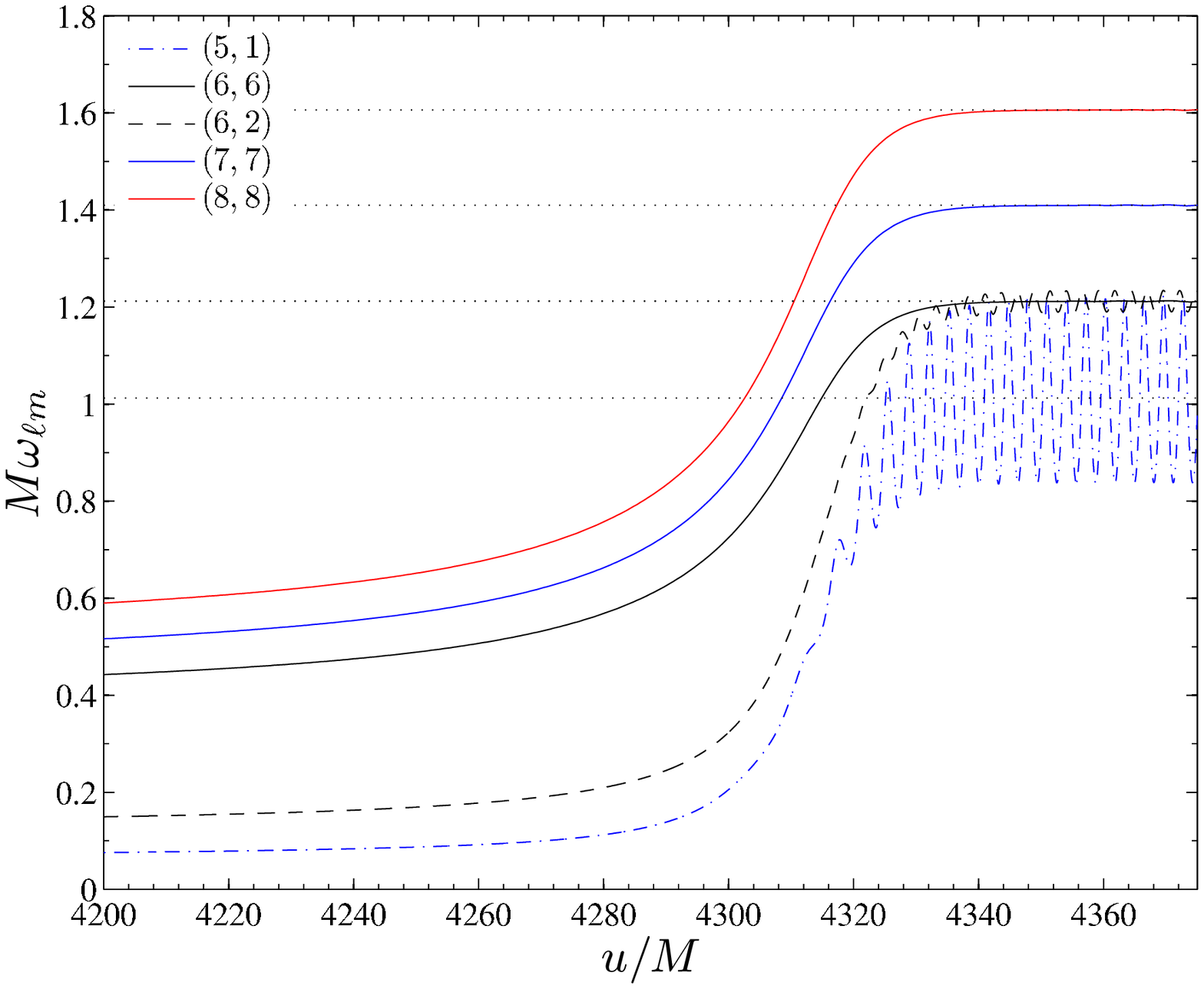}\\
  \caption{\label{fig:module_multipoles}Multipolar structure of the waveform.
    The left panels exhibit the moduli; the right panels the instantaneous
    gravitational wave frequencies for some representative multipoles.
    Note the oscillation pattern during ringdown (especially in the $\ell=2$,
    $m=1$ modulus and frequency) due to the interference between positve 
    and negative frequency QNMs. The waveform refer to the $\nu=10^{-3}$ mass ratio.}
\end{center} 
\end{figure*}

As for the instantaneous GW frequency, the right-panels 
of Fig.~\ref{fig:module_multipoles} show the same kind of 
behavior for each multipole:  $M\omega_{\ell m}$ is approximately
equal to $m\Omega$ during the inspiral, to grow abruptly during
the nonadiabatic plunge phase until it saturates at the ringdown frequency
(indicated by dashed lines in the plot). As already pointed out
in Ref.~\cite{Nagar:2006xv,Damour:2007xr} the oscillation pattern 
that is clearly visible for some multipoles is due to the contemporary
(but asymmetric) excitation of the positive and negative frequency
QNMs of the black hole. We shall give details on this phenomenon in
Sec.~\ref{sec:fitringdown}. 

\subsection{quasiuniversal plunge}
\label{sec:qgeoplunge}

In this section we discuss in quantitative terms the relative contribution 
of each multipole during the plunge, merger and ringdown phase. 
The analysis is based on the energy and angular momentum computed from the
emitted GW. While these quantities represent a ``synthesis'' of the
information we need, their computation and interpretation have some
subtle points that are discussed below. 

For a (adiabatic) sequence of circular orbits, this information
was originally obtained in Cutler et al.~\cite{Cutler:1993vq};
for the radial plunge of a particle initially at rest at infinity, 
the classical work of Davis, Ruffini, Press and Price~\cite{Davis:1971gg} 
found that about the $90\%$ of the total energy is quadrupole ($\ell=2$) 
radiation, and about the $8\%$ is octupole ($\ell=3$) radiation.
Concerning the transition from quasicircular inspiral to plunge,
Ref.~\cite{Nagar:2006xv} performed a (preliminary) calculation 
of the total  energy and angular momentum losses during a ``plunge'' 
phase (that was defined by the condition $r<5.9865M$, with $\nu=0.01$) 
followed by merger and ringdown, computing all the multipolar 
contributions up to  $\ell=4$ (see Table~2 in~\cite{Nagar:2006xv}).

We will follow up and improve the calculation of Ref.~\cite{Nagar:2006xv}.
Let us first point out some conceptual difficulties.
As a matter of fact, any kind of 
computation of the losses during the transition from
inspiral to plunge in our setup will depend both on the 
value of $\nu$, and on the initial time from which one starts the 
integration of the fluxes (for instance on the time when one 
defines the beginning of the ``plunge'' phase). 
It follows that, if robust and meaningful results are desired, the
calculation has to be focused on the part of the waveforms 
that is quasiuniversal (i.e., with
negligible dependence on $\nu$). As was pointed 
out in~\cite{Nagar:2006xv}, the quasiuniversal 
behavior reached in the $\nu\to 0$ limit is linked to the 
quasigeodesic character of the plunge motion, which 
approaches the geodesic which starts from the LSO in the 
infinite past with zero radial velocity.

In this respect, let us recall that, as shown in 
Ref.~\cite{Buonanno:2000ef}, the transition from the adiabatic 
inspiral to the nonadiabatic plunge is {\it not sharp}, but rather 
{\it blurred}, namely  it occurs in a radial domain around the LSO 
which scaled with $\nu$ as $r-6M\sim \alpha M\nu^{2/5}$, 
with the radial velocity scaling as $v_r\sim -\beta\nu^{3/5}$. 
In practical terms, this means that the quasiuniversal, quasigeodesic 
plunge does not really start at $r_0 = 6M$, but at 
about $r_0/M\sim 6-\alpha\nu^{2/5}$. 
In Ref.~\cite{Buonanno:2000ef}, using a 2.5PN Pad\'e resummed
radiation reaction, the coefficients $\alpha$ and $\beta$ were determined
to be $\alpha_{\rm 2.5PN}=1.89$ and $\beta_{\rm 2.5PN}=-0.072$. 
However, since our setup is based on the 5PN resummed radiation 
reaction force, we do not expect those numbers to remain unchanged, 
so that they do not represent for us a reliable estimate to extract 
the part of the waveforms we are interested in. 
Taking a pragmatical approach, we can determine this quasiuniversal 
region by contrasting our simulations at different $\nu$, so to 
see when the dependence on $\nu$ is 
sufficiently ``small'' (say at $1\%$ level in the energy 
and angular momentum losses, see below).

Figure~\ref{fig:conv_with_mu},  displays the ``convergence''
to the $\ell=m=2$ modulus (upper panel) and frequency (lower panel)
for the three values of $\nu$. For convenience the waveforms
have been time-shifted so that the maxima of the waveform 
mudulus (located at  $u-u_{\rm max}=0$ in the figure) 
coincide. The plot clearly shows that the late-time part 
of the waveform has a converging trend to some 
``universal'' pattern that progressively approximates  
the ``exact'' $\nu=0$ case. Note that, at the visual level, 
amplitudes and frequencies for  $\nu=\{10^{-3}\,10^{-4}\}$ 
look barely distinguishable during the late part of 
the plunge, entailing a very weak dependence on the 
properties of radiation reaction.  From this analysis we 
can assume a quasiuniversal and quasigeodesic plunge 
starting at about $u-u_{\rm max}=-50$ (vertical dashed line),  
which corresponds to  $M\omega_{22}\simeq 0.167$, which 
is about $1.23\times (2\Omega_{\rm LSO})$ (for reference,
we indicate with a horizontal line the $2\Omega_{\rm LSO}$
frequency in the lower panel of the figure)~\footnote{Note that 
the radial separation that corresponds to $u-u_{\rm max}=-50$ 
is $r\simeq 5.2M$ [more precisely, $r\simeq 5.199M$ $(5.228M$) 
for $\nu=10^{-4}$ ($\nu=10^{-3}$)], i.e., we have a $13\%$ 
difference with  the value, $5.88$ obtained using the former 
EOB analysis with $\alpha=1.89$.}.
We integrate the multipolar energy and angular momentum 
fluxes from $u-u_{\rm max}=-50M$ onwards and sum over all the multipoles
up to $\ell=8$. The outcome of this computation is listed
in Table~\ref{tab:table1} for the $\nu=\{10^{-3},\, 10^{-4}\}$.
Note that the agreement of these numbers at the level of $1\%$ 
is a good indication of the quasigeodesic character of the
dynamics behind the part of the waveform that we have selected.
The numerical information of Table~\ref{tab:table1} is 
completed by Tables~\ref{tab_app1}-\ref{tab_app2}
in Appendix~\ref{sec:losses}, were we list the values and
the relative weight of each partial multipolar contribution.
Coming thus to the main conclusion of this analysis, it turns out 
that the $\ell=m=2$ multipole contributes to the total 
energy (angular momentum) for about the $58\%$ ($62\%$), 
the $\ell=m=3$ for about the $20\%$ ($20\%$) , the $\ell=m=4$ for
about the $8\%$ ($7.6\%$) and the $\ell=m=5$ for the $3.5\%$ ($3.3\%$). 
For what concerns the odd-parity multipole, the dominant
one, $\ell=2$, $m=1$, contributes to $4.3\%$ of the total energy 
and $2.3\%$ of the total angular momentum.
We address again the reader to Appendix~\ref{sec:losses} for the fully
precise quantitative information.

\begin{figure}[t]
\begin{center}
  \includegraphics[width=0.45\textwidth]{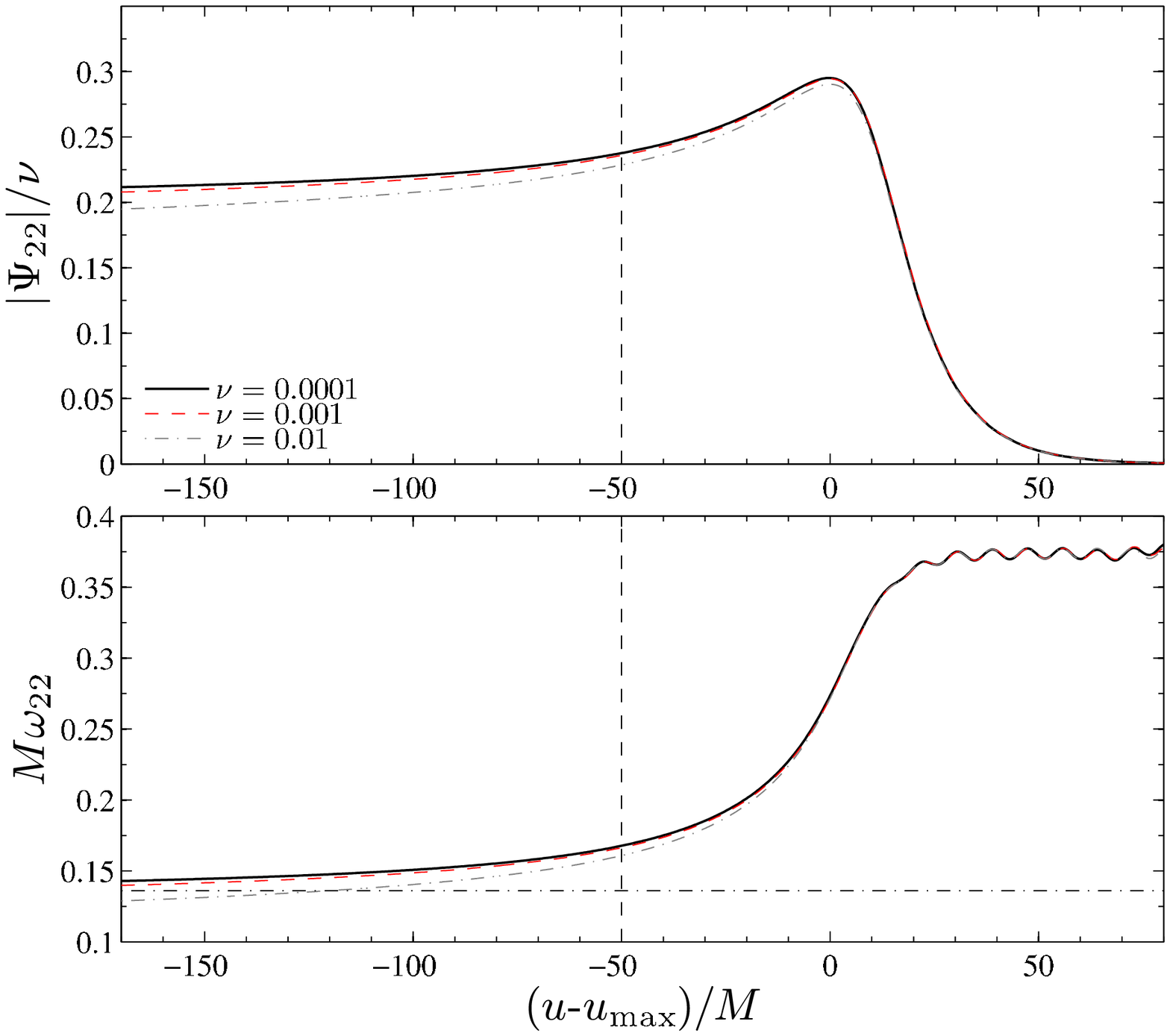}
  \caption{\label{fig:conv_with_mu} ``Convergence'' of the waveform 
    when $\nu\to 0$. Retarded times have been shifted so that the 
    zero coincides with the maximum of the waveform modulus 
    $|\Psi^{(\rm e)}_{22}|$ for each value of $\nu$. The horizontal
    dashed line indicates the adiabatic LSO frequency. The vertical dashed 
    line conventionally identifies the beginning of an approximately 
    quasiuniversal and quasigeodesic plunge phase.}
\end{center} 
\end{figure}

\subsection{Ringdown}
\label{sec:fitringdown}

Let us focus now on the analysis of the waveform during pure ringdown only. 
Our main aim here is to extract quantitative information from the
oscillations that are apparent in the gravitational wave frequency (and
modulus) during ringdown (see Fig.~\ref{fig:module_multipoles}). 
As explained in Sec.~IIIB of 
Ref.~\cite{Damour:2007xr}, the physical interpretation 
of this phenomenon is clear, namely it is due to an asymmetric 
excitation of the positive and negative QNM frequencies of the black hole 
triggered by the ``sign'' of the particle motion (clockwise or counterclockwise). 
The modes that have the {\it same sign} of $m\Omega$ are the dominant
ones, while the others with opposite sign are less excited (smaller amplitude).
Since QNMs are basically excited by a resonance mechanism, their
strength (amplitude) for a given multipole $(\ell,m)$ depends 
on their ``distance'' to the critical (real) exciting frequency
$m\Omega_{\rm max}$ of the source, where $\Omega_{\rm max}$ indicates
the maximum of the orbital frequency.
In our setup, the particle is inspiralling counterclockwise 
(i.e., $\Omega >0$), therefore the positive frequency QNMs 
are more excited than the negative frequency ones. 
The amount of (relative) excitation will depend on $m$.
Such QNM ``interference'' phenomenon was noted and explained already
in Refs.~\cite{Nagar:2006xv,Damour:2007xr}, although no quantitative
information was actually extracted from the numerical data. We 
perform here this quantitative analysis.

\begin{table}[t]
\caption{\label{tab:table1}Total energy and angular momentum emitted
during the quasiuniversal, quasigeodesic plunge phase, the merger and 
ringdown (it is defined by the condition 
$M\omega_{22}\gtrsim 0.167$, see Fig.~\ref{fig:conv_with_mu} ).}
 \begin{center}
  \begin{ruledtabular}
  \begin{tabular}{lcc}
    $\nu$ & $M\Delta E/\mu^2$  & $\Delta J/\mu^2$ \\
   \hline \hline 
 $10^{-3}$     & 0.47688  &  3.48918 \\
 $10^{-4}$    & 0.47097  &  3.44271 \\
 \end{tabular}
\end{ruledtabular}
\end{center}
\end{table}

The waveform during the ringdown has the structure
\be
\label{eq:qnm}
\Psi_{\lm}^{(\rm e/o)}=\sum_n C_{\ell m n}^+ e^{-\sigma^+_{\ell n}t} + \sum_n
C_{\ell m n}^- e^{-\sigma^-_{\ell n}t}, 
\ee
were we use the notation of Refs.~\cite{Damour:2006tr,Damour:2007xr},
and denote the QNM complex frequencies with
$\sigma_{\ell n}^{\pm}=\alpha_{\ell n}\pm \omega_{\ell n}$
and $C^{\pm}_{\ell m n}$ the corresponding complex amplitudes.
For each value of $\ell$, $n$ indicates the order of the mode, 
$\alpha_{\ell n}$ its inverse damping time and $\omega_{\ell n}$ its 
frequency. For example, 
defining $a_{\ell m n}e^{\ii {\vartheta}_{\ell m n}}\equiv C^{-}_{\ell m n}/C^+_{\ell m n}$, 
in the presence of only one QNM (e.g., the fundamental one, $n=0$) 
the instantaneous frequency computed from Eq.~(\ref{eq:qnm}) reads
\bea
\label{eq:qnm:interference}
\omega_{\lm}^{(\rm e/o)}&=& 
-\Im\left(\frac{\dot{\Psi}_{\ell m}^{(\rm e/o)}}{\Psi_{\ell m}^{(\rm e/o)}}\right) \\
&=& \frac{\left(1-a_{\ell m 0}^2\right)\omega_{\ell 0}}
      {1+a_{\ell m 0}^2+2a_{\ell m 0}\cos\left(2 \omega_{\ell 0} t +
	\vartheta_{\ell m 0}\right)}\nonumber .
\eea
This simple formula illustrates that, if the two modes are equally excited 
($a_{\ell 0}=1$) then there is a destructive interference and the
instantaneous frequency is zero; on the contrary, if one mode (say the
positive one) is more excited than the other, the instantaneous frequency 
oscillates around a constant value that asymptotically tends to 
$\omega^+_{\ell 0}$ when $C_{\ell m 0}^-\rightarrow0$. 
In general, one can use a more sofisticated version of
Eq.~\eqref{eq:qnm:interference}, that includes various overtones for 
a given multipolar order, as a template to fit the instantaneous GW frequency 
and to measure the various $a_{\ell m n}$ and  $\vartheta_{\ell m n}$ 
during the ringdown. 
For simplicity, we concentrate here only on the measure of  $a_{\ell m 0}$,
and we use directly Eq.~\eqref{eq:qnm:interference}. To perform such
a fit\footnote{For 
this particular investigation we use $\nu=10^{-2}$ data. The reason 
for this choice is that, in our grid setup, the waveforms are practically 
causally disconnected by the boundaries and we have a longer and cleaner
ringdown than in the other two cases.} 
(with a least-square method) we consider only the part of the
ringdown that is dominated by the fundamental (least-damped) QNM; i.e., 
the ``plateau of oscillations'' approximately starting at $u/M= 4340$ 
in the right-panels of Fig.~\ref{fig:module_multipoles}.
\begin{table}[t]
  \caption{\label{tab:rngdwn:fit} Fit of QNM interference with Eq.~(\ref{eq:qnm:interference}) 
    for a representative sample of multipoles. 
    The numbers refer to $\nu=10^{-2}$. The $M\omega_{\ell 0}$ 
    column lists the values of the fundamental QNMs frequencies gathered from 
    the literature~\cite{Chandrasekhar:1975zza,Chandrasekhar:1985kt,
    Leaver:1985ax,Berti:2005ys} (see also Ref.~\cite{Berti:2009kk} for a recent 
    review and for highly accurate computations).
    By contrast, the primed values are obtained from our numerical 
    data by fitting the ringdown frequency for \emph{both} $a_{\ell m 0}$ and 
    $M\omega_{\ell0}$. Note the good consistency between the two methods.}
 \begin{center}
  \begin{ruledtabular}
  \begin{tabular}{llcccc}
    $\ell$ & $m$ & $a_{\ell m 0}$ & $a'_{\ell m 0}$  & $M\omega_{\ell0}$ & $M\omega'_{\ell0}$\\
   \hline \hline
   2 & 1 & 7.2672$\times10^{-2}$   & 7.2678$\times10^{-2}$ & 0.37367   &  0.37369  \\
   2 & 2 & 4.8476$\times10^{-3}$   & 4.848$\times10^{-3}$  & 0.37367   &  0.37361   \\
   3 & 1 & 9.3403$\times10^{-2}$   & 9.3403$\times10^{-2}$ & 0.59944   &  0.59944  \\
   3 & 2 & 8.008$\times10^{-3}$    & $8.011\times10^{-3}$  & 0.59944   &  0.59936  \\
   3 & 3 & 5.5471$\times10^{-4}$   & 5.5477$\times10^{-4}$ & 0.59944   &  0.59943  \\
   4 & 1 & 9.1560$\times10^{-2}$   & 9.1559$\times10^{-2}$ & 0.80917   &  0.80918  \\
   4 & 2 & 9.1433$\times10^{-3}$   & 9.1435$\times10^{-3}$ & 0.80917   &  0.80917  \\
   4 & 3 & 9.0473$\times10^{-4}$   & 9.0475$\times10^{-4}$ & 0.80917   &  0.80917  \\
   4 & 4 & 6.382$\times10^{-5}$    &  6.379$\times10^{-5}$ & 0.80917   &  0.80918   \\
  \end{tabular}
\end{ruledtabular}
\end{center}
\end{table}
The fundamental frequency $n=0$ has been used as given input, 
and we fit for the amplitude ratio $a_{\ell m 0}$ and relative 
phase $\theta_{\ell m 0}$. The outcome of the fit for some multipoles 
is exhibited in Table~\ref{tab:rngdwn:fit}. Note that in the third
and fourth column we list also the values that one obtains by fitting
{\it also} for the frequency $\omega_{\ell 0}'$. We obtain perfectly
consistent results.
Note that for the multipole $\ell=8$ we were obliged to compute 
the frequency only in this way, since we could not find this number in the results 
of~\cite{Berti:2009kk}:  we obtain the value $M\omega_{8 0}=1.60619$. 
The table quantifies that the strongest interference pattern, that always
occurs for $m=1$ (for any $\ell$), corresponds to a relative contribution of
the negative frequency mode of the order of about $9\%$. 
This trend remains true for all values of $\ell$. For example, 
we have $a_{810}=9.54\times 10^{-2}$ and $a_{710}=9.48\times 10^{-2}$. 
Note finally that the presence of the negative mode for the $\ell=2$, $m=1$
shows up also in the corresponding modulus $|\Psi_{21}|/\nu$, with the
characteristic oscillating pattern superposed to the 
exponential decay (see top-left panel of Fig.~\ref{fig:module_multipoles}).
[See also Ref.~\cite{Hadar:2009ip} for an analytical treatment 
of the ringdown excitation amplitudes during the plunge].

\section{Gravitational recoil}
\label{sec:recoil}

Let us now come to the computation of the gravitational 
recoil, or ``kick'', imparted to the system due to the 
anisotropic emission of gravitational radiation. 
The calculation of these kicks in general relativity 
has been carried out in a variety of ways, that before 2005
relied mainly on analytical and semianalytical techniques. 
In particular, let us mention that, after the pioneering 
calculation of Fitchett~\cite{Fitchett:1983} and 
Fitchett and Detweiler~\cite{Fitchett:1984qn}, earlier estimates 
included a perturbative calculation~\cite{Favata:2004wz}, 
a close-limit calculation~\cite{Sopuerta:2006wj} 
and a post-Newtonian calculation valid during 
the inspiral phase only~\cite{Blanchet:2005rj}. 
This latter calculation has been recently improved 
by bringing together post-Newtonian theory and the 
close limit approximation, yielding close agreement
with purely NR results~\cite{LeTiec:2009yg}.
In addition, a first attempt to compute the final 
kick within the EOB approach~\cite{Damour:2006tr} yielded 
the {\it analytical understanding} (before any numerical result
were available) of the qualitative behavior of the kick 
velocity (notably the so-called ``antikick'' phase), 
as driven by the intrinsically nonadiabatic character 
of the plunge phase. Such preliminary EOB 
calculation was then improved in~\cite{Schnittman:2007sn}, 
which included also inputs from NR simulations.
On the numerical side, after the pioneering computation of 
Baker et al.~\cite{Baker:2006vn}, there has been a 
plethora of computations of the kick from spinning black holes 
binaries, focusing in particular on  the so-called superkick
configurations. By contrast, for the nonspinining case,
Refs.~\cite{Gonzalez:2006md,Gonzalez:2008bi} represent
to date the largest span of mass ratios for which 
the final kick velocity is known (see also Ref.~\cite{Pollney:2007ss} 
for the nonprecessing, equal-mass spinning case).
In addition, the use of semianalytical models prompted a deeper
understanding of the structure of the gravitational recoil 
as computed in NR simulations~\cite{Schnittman:2007ij}.
However, despite all these numerical efforts, to date  
there are no ``numerical'' computations of the final recoil 
velocity  in the $\nu\to 0$ limit: the only estimates 
rely on fits to NR data of the form
\be 
\label{eq:kick}
v^{\rm kick}=A\nu^2\sqrt{1-4\nu}(1+B\nu )\ ,
\ee
with the coefficient $A$ giving the extrapolated  value in 
the $\nu\to 0$ limit~\cite{Gonzalez:2006md,Gonzalez:2008bi}. 
The aim of this section is to provide a value of $A$ that 
comes from an actual (numerical) computation within 
perturbation theory. 

It is convenient to treat the kick velocity 
vector imparted to the system by GW emission as a complex 
quantity, i.e. 
$v\equiv v_x+\ii v_y$.
By integrating Eq.~\eqref{eq:dPdt} in time and by changing the
sign, the (complex) velocity accumulated by the system up to
a certain time $t$ is given by
\be
\label{eq:kick_one}
v \equiv v_x + {\rm i}v_y= -\dfrac{1}{M}\int_{-\infty}^t\left(\F_x^{\bf P} + {\rm i}\F_y^{\bf P}\right) dt' .
\ee
Since in practical situations one is always dealing with a 
finite timeseries for the linear momentum flux, it is not 
possible to begin the integration from $t=-\infty$, but rather
at a finite initial time $t_0$. This then amounts in the need
of fixing some (vectorial) integration constant $v_0$
that accounts for the velocity that the system has acquired in
evolving from $t=-\infty$ to $t=t_0$, i.e.
\be
\label{eq:kick_two}
v = v_0 -\dfrac{1}{M}\int_{t_0}^t\left(\F_x^{\bf P} + {\rm i}\F_y^{\bf P}\right) dt.
\ee
As it was emphasized in Ref.~\cite{Pollney:2007ss}, the proper
inclusion of $v_0$ is crucial to get the correct (monotonic) 
qualitative and quantitative behavior of the time evolution 
of the magnitude $|v|$ of the recoil velocity. 
Typically, not only the final value 
of $|v|$ may be wrong of about a $10\%$, but one can also 
have spurious oscillations in  $|v|$ during the inspiral phase
if $v_0$ is not properly determined or simply set to zero. 
See in this respect Sec.~IVA of Ref.~\cite{Pollney:2007ss}.
\begin{figure}[t]
\begin{center}
  \includegraphics[width=0.45\textwidth]{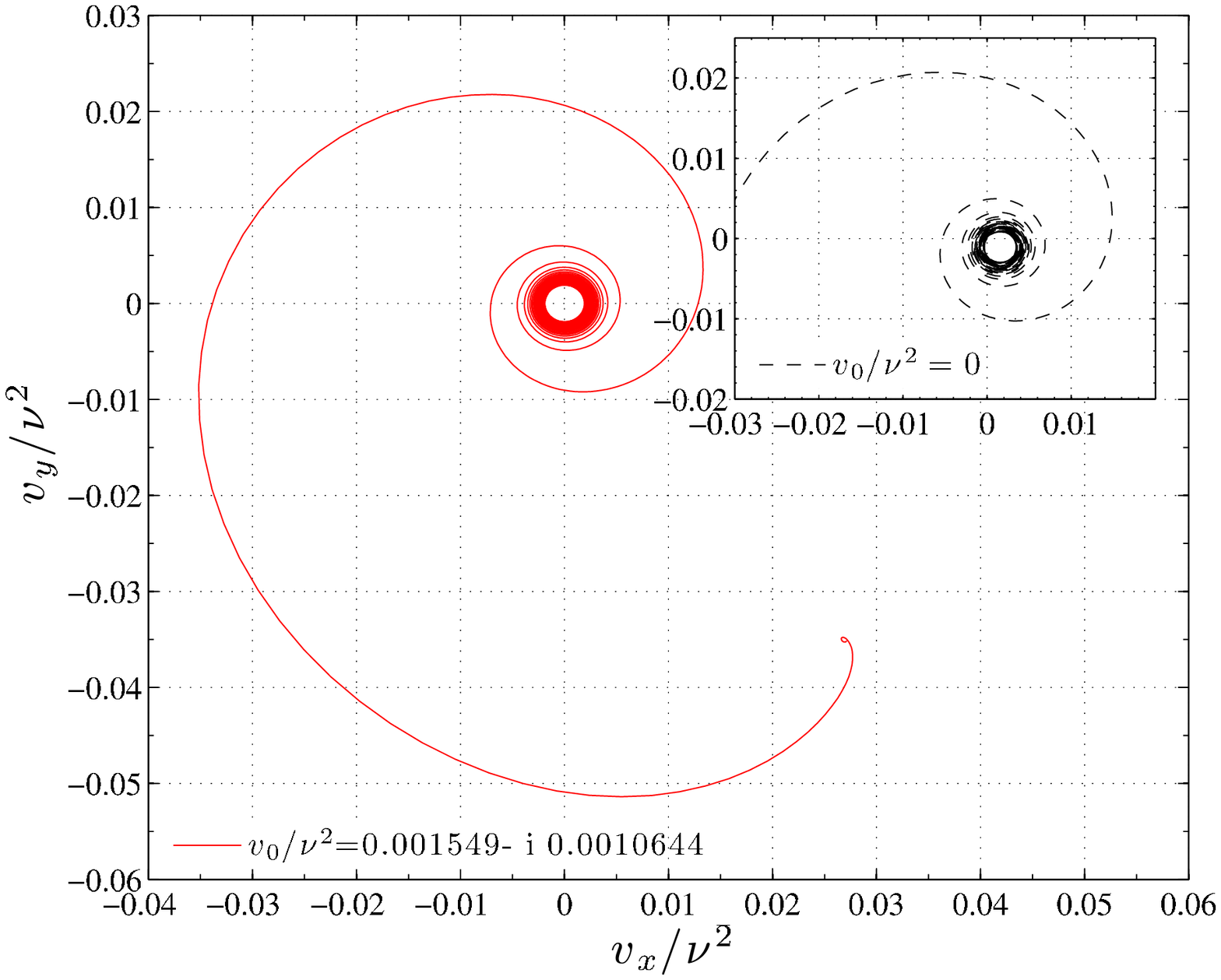}
  \caption{\label{fig:hodograph}Parametric plot of 
   $v_x$ versus $v_y$ (for $\nu=10^{-3}$) obtained from 
   Eq.~\eqref{eq:kick_two} with 
   $v_0/\nu^2 = (1.549-\ii 1.0644)\times 10^{-3}$. 
   The analogous plot with $v_0=0$ is shown in the inset.}
\end{center} 
\end{figure}

As in Ref.~\cite{Pollney:2007ss}, the numerical determination 
of $v_0$ can be done with the help of the ``hodograph'', i.e., 
a parametric plot of the instantaneous velocity vector in the 
complex velocity plane $(v_x,v_y)$. This hodograph is displayed 
in Fig.~\ref{fig:hodograph} for $\nu=10^{-3}$. 
Let us focus first on the inset, that exhibits the outcome 
of the time integration with $v_0=0$. 
Note that the center of the inspiral (corresponding to the 
velocity accumulated during the quasiadiabatic inspiral phase) 
is displaced with respect to the correct value  $v=(0,0)$,
corresponding to $v_0=0$ at $t=-\infty$.
The initial $v_x^0$ and $v_y^0$ are determined 
as the translational ``shifts'' that one needs to add 
(in both $v_x$ and $v_y$) so that   the ``center'' of this 
inspiral is approximately zero. The result of this 
operation led (for $\nu=10^{-3}$) to 
$v_x^0/\nu^2=1.1549\times 10^{-3}$ and $v_y^0/\nu^2=-1.0644\times
10^{-3}$; this is displayed in the main panel of Fig.~\ref{fig:hodograph}.

\begin{figure}[t]
\begin{center}
  \includegraphics[width=0.45\textwidth]{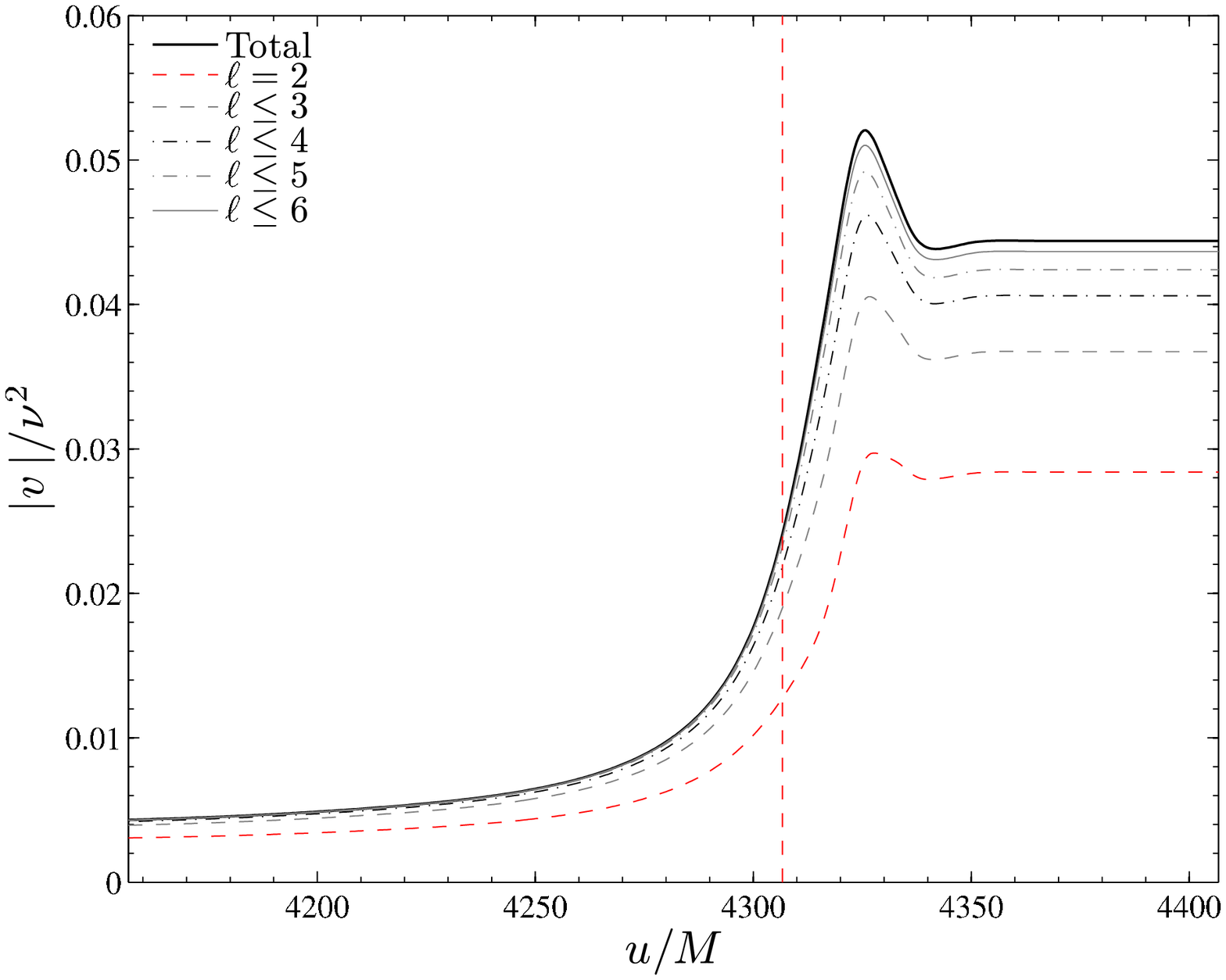} 
  \caption{\label{fig:recoil}Time-evolution of the magnitude of the recoil
  velocity. The figure shows the monotonic ``multipolar'' convergence
  to the final result. The plot refers to mass ratio $\nu=10^{-3}$.}
\end{center} 
\end{figure}

This judicious choice of the integration constant is such that 
the modulus $|v|=\sqrt{v_x^2 + v_y^2}$ of the  accumulated recoil 
velocity grows essentially monotonically in time and no spurious
oscillations are present during the inspiral phase. This is 
emphasized by  Fig.~\ref{fig:recoil}. In the figure, we show,
as a solid line, the modulus of the {\it total} accumulated 
kick velocity  versus observer's retarded time (as before, waveforms
are extracted at $r_*^{\rm obs}/M=1000$). This ``global'' computation 
is done including in the sum of Eq.~\eqref{eq:dPdt} all the partial
multipolar contribution up to $\ell=7$ (which actually means 
considering also the interference terms between 
$\ell=7$ and $\ell=8$ modes). To guide the eye, we added 
a vertical dashed line locating the maximum of 
$|\Psi^{(\rm e)}_{22}|$, that approximately corresponds to 
the dynamical time when the particle crosses the light-ring. 
Note the typical shape of $|v|$, with a clean local 
maximum and the so-called ``antikick'' behavior, that is 
qualitatively identical to the corresponding curves 
computed (for different mass ratios) by NR simulations
(see for example Fig.~1 of Ref.~\cite{Schnittman:2007ij}
for the 2:1 mass ratio case).

In addition to the total  recoil magnitude computed up to $\ell=7$,
we display on the same plot also the ``partial'' contribution, i.e.
computations of $|v|$ where we truncate the sum over 
$\ell$ in Eq.~\eqref{eq:dPdt} 
at a given value $\ell^*<7$. In the figure we show
(depicted as various type of nonsolid lines) the evolution of 
recoil with $2\leq\ell^*\leq 6$. 
Note that each partial-$\ell$ contribution to the linear momentum 
flux has been integrated in time (with the
related choice of integration constants) before performing the 
vectorial sum to obtain the total $v$. The fact that each 
curve nicely grows monotonically without spurious oscillations
during the late-inspiral phase is a convincing indication of the 
robustness of the procedure we used to determine $v_0^{\ell^*}$
by means of hodographs\footnote{Note that the procedure can actually 
be automatized by determining the ``baricenter'' of the adiabatic 
inspiral in the $(v_x,v_y)$ plane corresponding to early evolution.}.
\begin{table}[t]
\caption{\label{tab:recoil}Magnitude of the final and maximum kick
  velocities for the three values of $\nu$ considered. The last row 
  lists the values extrapolated to $\nu=0$ from  $\nu=\{10^{-3},10^{-4}\}$ data.}
 \begin{center}
  \begin{ruledtabular}
  \begin{tabular}{lcc}
    $\nu$ &  $|v^{\rm end}|/\nu^2$ & $|v^{\rm max}|/\nu^2$ \\
   \hline \hline
    $10^{-2}$   &  0.043234     &  0.050547  \\
    $10^{-3}$   &  0.044401     &  0.052058  \\
    $10^{-4}$   &  0.044587     &  0.052298  \\
   \hline 
   {\bf 0}    &   {\bf 0.0446} & {\bf 0.0523}  \\
  \end{tabular}
\end{ruledtabular}
\end{center}
\end{table}
The figure highlights at a visual level the influence of
high multipoles to achieve an accurate results. To give some 
meaningful numbers, if we consider $\ell^*=6$, we obtain 
$|v_6^{\rm fin}|/\nu^2=0.0437$ (i.e. $1.7\%$ difference), 
while  $|v_5^{\rm fin}|/\nu^2= 0.0424$ ($4.5\%$ difference).

The information conveyed by Fig.~\ref{fig:recoil} is completed by
Table~\ref{tab:recoil}, where we list the final value of the
modulus of the  recoil velocity of the center of mass $|v^{\rm end}|/\nu^2$
(as well as the corresponding maximum value $|v^{\rm max}|/\nu^2$)
obtained in our setup for the three values of $\nu$ that we have
considered. The computation of the kick for the other values of $\nu$ 
is procedurally identical and thus we show only the final numbers.
The good agreement between the three numbers is consistent with
the interpretation that the recoil is almost completely determined 
by the nonadiabatic plunge  phase of the system (as emphasized 
in Ref.~\cite{Damour:2006tr}), and thus it is almost unaffected 
by the details of the inspiral phase. Because of the late-plunge
consistency between waveforms that we showed above 
for $\nu=\{10^{-3},\,10^{-4}\}$, we have decided to extrapolate the
corresponding values of the kick for $\nu=0$. The corresponding
numbers are listed (in bold) in the last row of Table~\ref{tab:recoil}.

\begin{table}[t]
\caption{\label{tab:other_recoil}Recoil velocities in the test-mass limit
 obtained by (extrapolating) different finite-mass results. 
 Our ``best'' value is shown in bold. See text for explanations.}
 \begin{center}
  \begin{ruledtabular}
  \begin{tabular}{lcc}
    Reference                             &  $|v^{\rm end}|/\nu^2$ &  \\
   \hline \hline
   Gonz\'alez {\it et al.}~\cite{Gonzalez:2006md}  &  0.04         &   \\
   \hline
   Damour and Gopakumar~\cite{Damour:2006tr}   &  [0.010,\,0.035]  &\\
   Schnittman and Buonanno~\cite{Schnittman:2007sn}   &  [0.018,\,0.041]  &\\
   Sopuerta {\it et al.}~\cite{Sopuerta:2006wj}  & [0.023,\,0.046] &  \\
   Le Tiec, Blanchet and Will~\cite{LeTiec:2009yg}      &  0.032        & \\
   \hline
   This work          &  {\bf 0.0446}  & \\
  \end{tabular}
\end{ruledtabular}
\end{center}
\end{table}

In Table~\ref{tab:other_recoil} we compare the value of the 
final recoil with that (extrapolated to the test-mass limit) 
obtained from NR simulations~\cite{Gonzalez:2006md} and with 
semianalytical or seminumerical predictions, 
like the EOB~\cite{Damour:2006tr,Schnittman:2007sn}, 
the close-limit approximation~\cite{Sopuerta:2006wj} 
(that all give a range, with rather large error bars)
and the recent calculation of Le Tiec et al.~\cite{LeTiec:2009yg}
based on a hybrid post-Newtonian-close-limit calculation

We conclude this section by discussing in more detail the comparison 
of our result with the NR-extrapolated value. Since the NR-extrapolated
value that we list in Table~\ref{tab:other_recoil} was obtained using
{\it only} the data of Ref.~\cite{Gonzalez:2006md} (without the 10:1
mass ratio simulation of~\cite{Gonzalez:2008bi}), we have decided to 
redo the fit with all the NR data together (that have been kindly given to
us by the Authors).
To improve the sensitivity of the fit when $\nu$ gets small, we first
factor out the $\nu^2$ dependence in the data (i.e., we consider 
$v^{\rm NR}/\nu^2$, by continuity with the test-mass result).
We then fit the data with the function   
\begin{equation}
\label{kickfit}
\tilde{f}(\nu) = A \sqrt{1-4\nu}\left( 1+ B \nu \right) \ .
\end{equation}
Table~\ref{tab:kickfit} displays the results of the fit obtained using: 
the NR data of Ref.~\cite{Gonzalez:2006md} (consistent with the published result),
first row; the joined information of
Refs.~\cite{Gonzalez:2006md,Gonzalez:2008bi}, 
second row; and the NR data of~\cite{Gonzalez:2006md,Gonzalez:2008bi} together
with the test-mass result calculated in this paper.  Note that the NR fit are 
perfectly consistent with the test-mass value: in particular, our extrapolated 
value $|v^{\rm end}|/\nu^2=0.0446$ shows an agreement of $1.5\%$ with the 
value of $A$ obtained from the fit to the most complete NR information
(in bold in Table~\ref{tab:kickfit}).

\begin{figure}[t]
\begin{center}
  \includegraphics[width=0.45\textwidth]{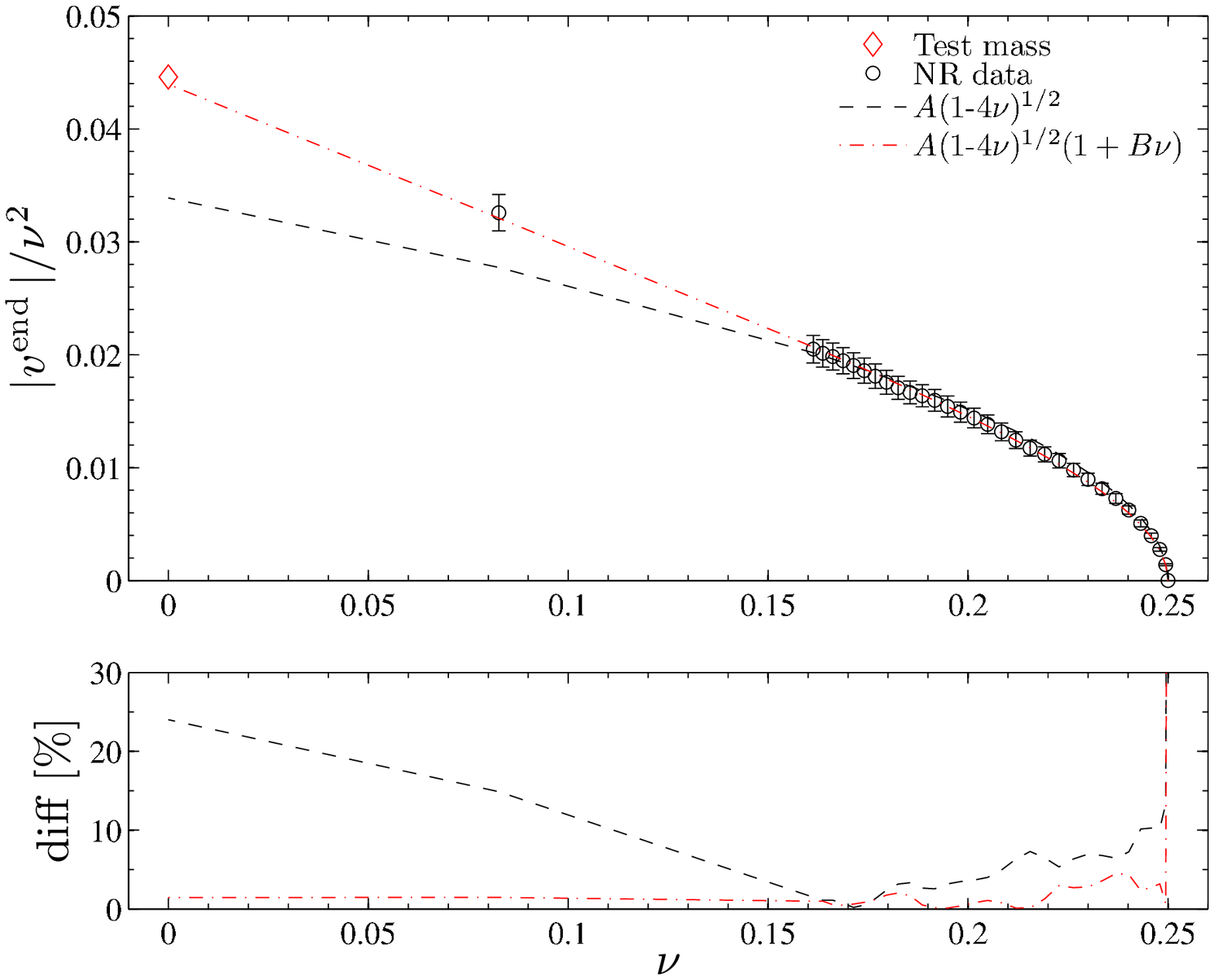} 
  \caption{\label{fig:fick} Results of the fit of 
   NR data of Refs.~\cite{Gonzalez:2006md,Gonzalez:2008bi} using 
   Eq.~\eqref{kickfit}. Note the good agreement between the NR-extrapolation
   and our test-mass result. The bottom panel contains the relative 
   difference with the data. This plot corresponds to the second row
   of Table~\ref{tab:kickfit}, without the test-mass point. See text for details.}
\end{center} 
\end{figure}

The information of the table is completed by Fig.~\ref{fig:fick}, that 
displays $\tilde{f}(\nu)$ (as a dash-dot line)  obtained from the complete 
NR data of Refs.~\cite{Gonzalez:2006md,Gonzalez:2008bi}. Note the visual 
good agreement between this extrapolation and the test-mass point when $\nu\to 0$.
For contrast, we also show on the plot (as a dashed line) the outcome of the 
fit with the simple Newtonian-like formula ($B=0$)~\cite{Fitchett:1983}.
We also tested the effect of adding a quadratic correction 
[i.e. a term $C\nu^2$ in the polynomial multiplying the square root 
in $\tilde{f}(\nu)$], but we found that it does not really improve 
the description of the data.
\begin{table}[t]
\caption{\label{tab:kickfit}Fit coefficients for the final magnitude of the kick 
  velocity from NR simulations as a function of $\nu$, Eq.~\eqref{kickfit}.
  See text for discussion.}
 \begin{center}
  \begin{ruledtabular}
  \begin{tabular}{lcc}
    Data & $A$  & $B$ \\
   \hline \hline
Gonz\'alez et al.~\cite{Gonzalez:2006md}                             &  0.04070  & -0.9883 \\
Gonz\'alez et al.~\cite{Gonzalez:2006md,Gonzalez:2008bi}            &  {\bf 0.04396}  & {\bf -1.3012}  \\
Gonz\'alez et al.~\cite{Gonzalez:2006md,Gonzalez:2008bi}+ This work &  0.04446  & -1.3482\\
\end{tabular}
\end{ruledtabular}
\end{center}
\end{table}

\section{Consistency checks}
\label{sec:checks}

In this section, we finally come to the discussion of some internal 
consistency checks of our approach. These consist in: 
(i) the verification of the consistency between the mechanical angular 
momentum loss (as driven by our analytical, resummed radiation 
reaction force) and the actual gravitational wave energy flux 
computed from the waves (as a follow up of a similar analysis 
done in Ref.~\cite{Damour:2007xr}); 
(ii) a brief analysis of the influence on the (quadrupolar) waveform
of the higher-order $\nu$-dependent EOB corrections entering 
the conservative  and nonconservative part of the relative dynamics.

\subsection{Angular momentum loss}

One of the results of Ref.~\cite{Damour:2007xr} was about 
the comparison between the mechanical angular momentum loss 
provided by the resummed radiation reaction $\hat{\F}_{\varphi}$ 
and the angular momentum flux computed from the multipolar waveform.
At that time, the main focus of Ref.~\cite{Damour:2007xr} was on 
to use of the ``exact'' instantaneous gravitational wave 
angular momentum flux $\dot{J}$ [see Eq.~\eqref{eq:dJdt}], 
to discriminate between two different expression of the
2.5PN Pad\'e resummed angular momentum flux 
$\F_\varphi^{\rm 2.5PN}$ that are degenerate during 
the adiabatic early inspiral.
In addition , in that setup it was also possible to:
(i) check consistency between $\dot{J}$ and $-\F_\varphi$ 
during the inspiral and early plunge; (ii) argue that non-quasicircular
corrections in the radiation reaction are present to produce
a good agreement between the ``analytical'' and the exact 
angular momentum fluxes also during the plunge, almost up 
to merger and (iii) show that  the ``exact'' flux is practically 
insensitive to (any kind of) NQC corrections.
Since we are now using a new radiation reaction force with 
respect to  Ref.~\cite{Damour:2007xr}, it is interesting to
redo the comparison between the Regge-Wheeler-Zerilli ``exact'' 
flux and the ``analytical'' mechanical loss computed along the
relative dynamics.
\begin{figure}[t]
\begin{center}
\includegraphics[width=0.45\textwidth]{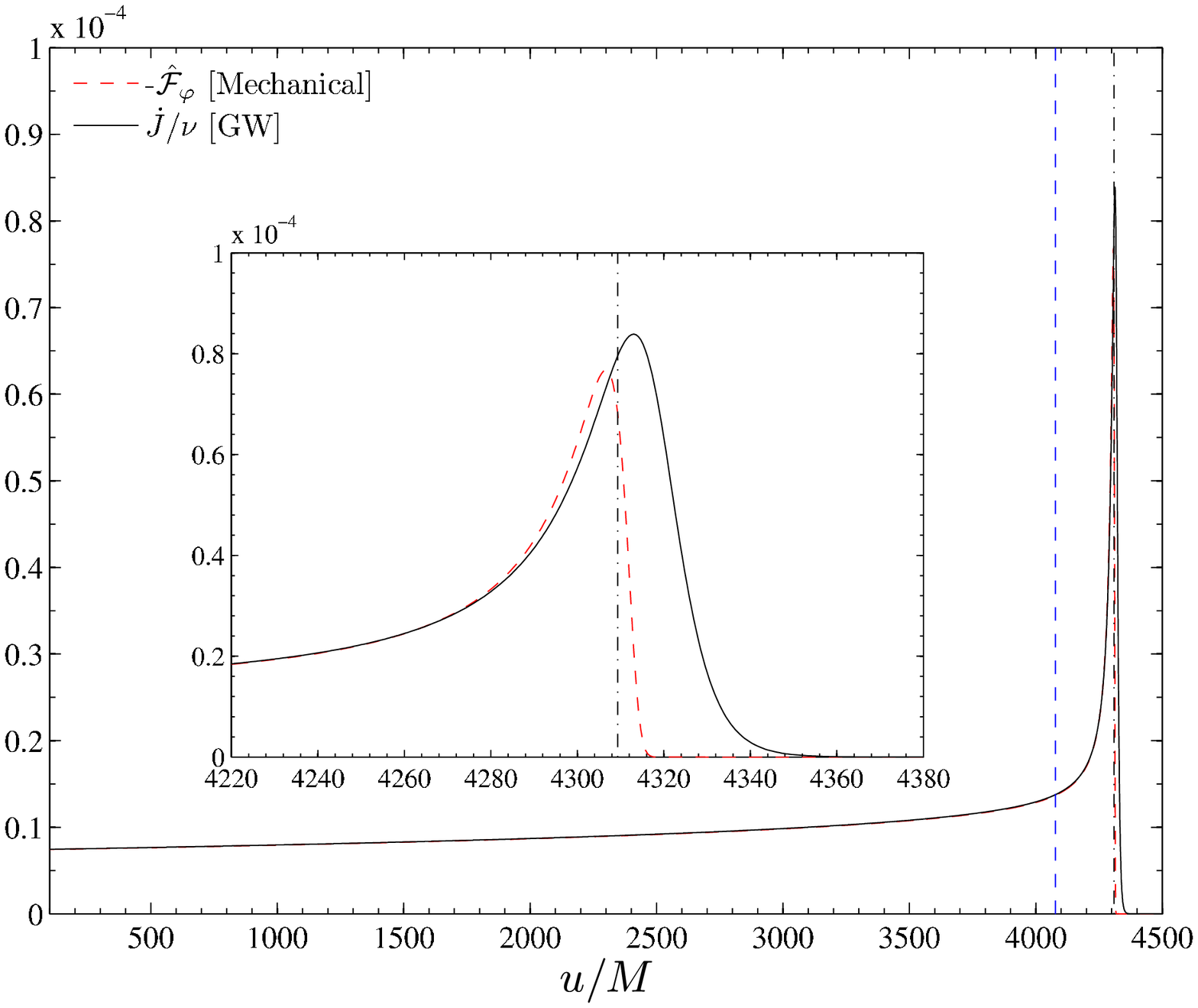}
\caption{\label{fig:dJdt_vs_time}Comparison between two angular 
momentum losses: the GW flux (solid line) computed \`a la Regge-Wheeler-Zerilli 
including up to $\ell=8$ radiation multipoles, and the mechanical angular momentum 
loss $-\F_\varphi$ (dash line). The two vertical lines correspond (from left to right) 
to the particle crossing respectively the adiabatic LSO location ($r=6M$) and 
the light-ring location ($r=3M$). The plot refers to $\nu=10^{-3}$.}
\end{center} 
\end{figure}
The result of this comparison is displayed in Fig.~\ref{fig:dJdt_vs_time},
that is the analogous of (part of) Fig.~2 of Ref.~\cite{Damour:2007xr}.
We show in the figure the mechanical angular momentum loss 
(changed of sign) $-\hat{\F}_\varphi/\nu$ versus the 
mechanical time $t/M$ together
with the instantaneous angular momentum flux $\dot{J}/\nu$
(computed from $\Psi^{(\rm e/o)}_{\lm}$ including all 
contributions up to $\ell=8$) versus observer's retarded time.
Note that we did not introduce here a possible shift between the
mechanical time $t$ and observer's retarded time $u$. As such
a shift is certainly expected to exist, our results should be
viewed as giving a lower bound on the agreement 
between $\F_\varphi$ and $\dot{J}$.
Note the very good visual agreement, not only above the LSO
(vertical dashed line) but also {\it below} the LSO,
and actually almost during the {\it entire} plunge phase. 
In fact, the accordance between the two fluxes is actually 
visually very good almost up to the merger 
(approximately identified by the maximum of the 
$\ell=m=2$ waveform, see dash-dot line 
in the inset)\footnote{Following the reasoning line of~\cite{Damour:2007xr},
the result displayed in the figure is telling us that most of the non-quasi
circular corrections to the waveforms (and energy flux) are already taken
into account automatically in our resummed flux, due to the intrinsic
dependence on it on $p_{r_*}$ through the Hamiltonian, so that one 
might need only to add pragmatically corrections that are very 
small in magnitude. These issues will deserve more careful 
investigations in forthcoming studies}.

\begin{figure}[t]
\begin{center}
\includegraphics[width=0.45\textwidth]{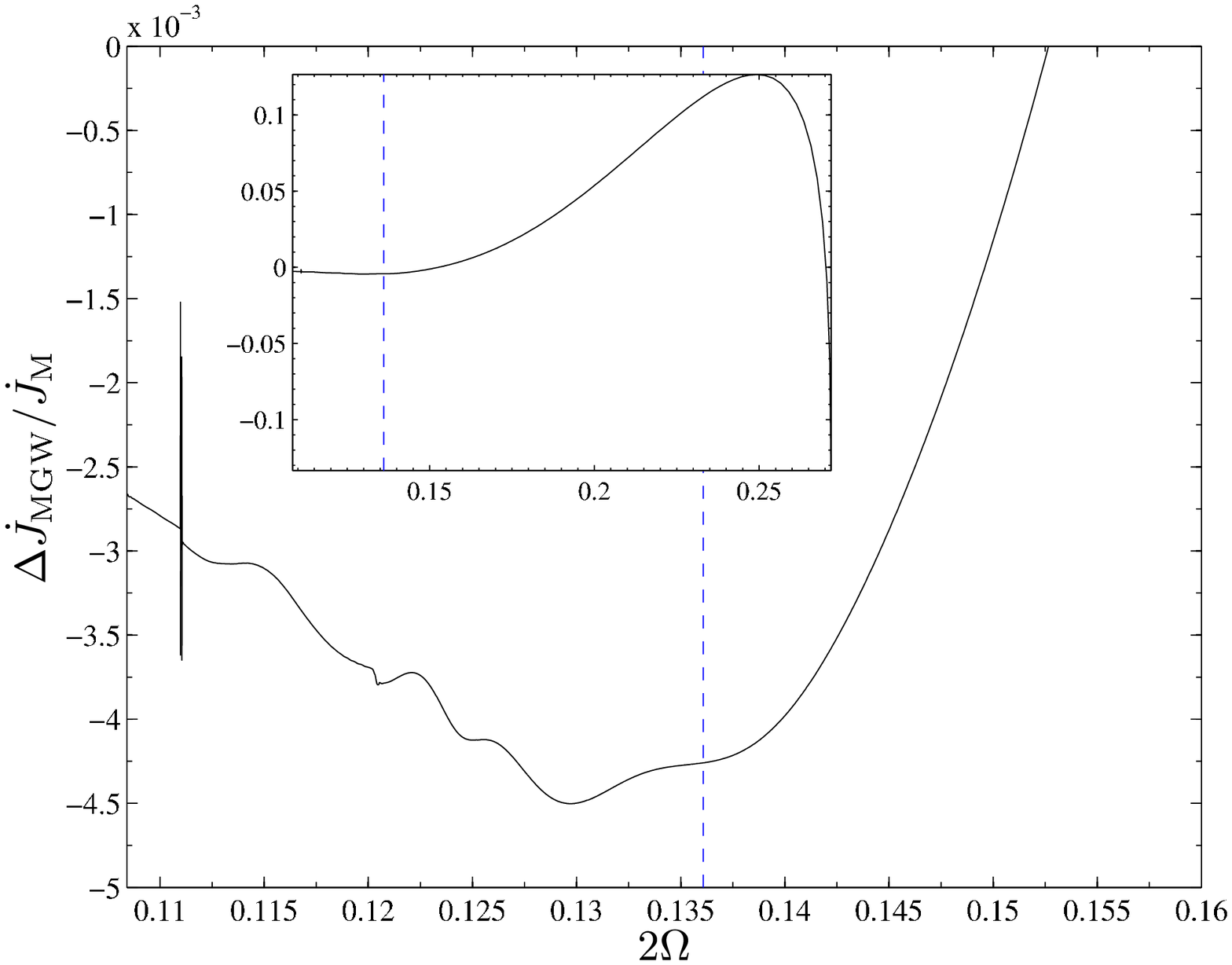}
\caption{\label{fig:DeltaJ}Difference between mechanical angular
  momentum loss and GW energy flux shown versus twice the orbital frequency
  $2\Omega$. The vertical line locates the adiabatic LSO frequency. The main
  panel focuses on the inspiral phase, while the inset shows the full range
  until the $2\Omega_{\rm max}$, where $\Omega_{\rm max}$ indicates the
  maximum of orbital frequency.}
\end{center} 
\end{figure}

We inspect this agreement at a more quantitative level in
Fig.~\ref{fig:DeltaJ}, where we plot the (relative) difference between 
$\dot{J}/\nu$ and $-\hat{\F}_\varphi/\nu$ versus (twice)
the orbital frequency. The inset shows the relative difference from
initial frequency to $2\Omega_{\rm max}$, where $\Omega_{\rm max}$  
is the maximum of the orbital frequency.
The main panel is a close-up centered around the LSO frequency.
Note that the relative difference is of the order of $10^{-3}$
during the late inspiral and the plunge, increasing at about only
a $10\%$ just before merger.
We have performed the same analysis for $\nu=10^{-2}$ and $\nu=10^{-4}$,
obtaining similar results. This is an indication that we have reached
the limit of accuracy of our resummation procedure, limit that evidently
is more apparent during the late part of the plunge. 
It is however remarkable that the fractional different 
is so small, confirming the validity of the
improved $\rho$-resummation of  Ref.~\cite{Damour:2008gu}.
In this respect, we mention in passing that this fractional difference 
can be made even smaller by further Pad\'e resumming the residual amplitude
corrections $\rho_{\ell m}$ in a proper way. This route was explored in
Ref.~\cite{Damour:2008gu} for the $\rho_{22}$ amplitude, yielding indeed
better agreement with the ``exact'' circularized waveform amplitude. 
A more detailed analysis of these delicate issues lies out of the 
purpose of this paper, but will be investigated in future work.

\subsection{Influence of dynamical ``self-force'' 
            $\nu$-dependent effects on the waveforms.}
\label{sec:eob_full}
In the work that we have presented so far we have included
in the relative dynamics only the leading order part of the
radiation reaction force, namely the one proportional to $\nu$.
This allowed us to compute, consistently as shown above, 
Regge-Wheeler-Zerilli-type waveforms. 
In doing so we have neglected all the finite-$\nu$ effects  
that are important in  the (complete) EOB description of the 
two-body problem, that is:  (i) $\nu$-dependent corrections to the
conservative part of the dynamics\footnote{These corrections 
come in both from the resummed EOB Hamiltonian  $H_{\rm EOB}$ with
the double-square-root structure and from the EOB radial 
potential $A(r,\nu)$.} and (ii) higher order $\nu$ dependent 
corrections in the nonconservative part of the dynamics, 
i.e. corrections entering in the definition of 
the angular momentum flux $\hat{\F}_{\varphi}$.

In this section we want to quantify the effects entailed by 
these corrections on our result. To do so, we switch on 
the ``self-force'' $\nu$-dependent corrections in the Hamiltonian
and in the flux defining the complete EOB relative dynamics and 
we compute EOB waveforms for $\nu=\{10^{-2},10^{-3},10^{-4}\}$.
Since this analysis aims at giving us only a general quantitative
idea of the effect of ``self-force'' corrections, we restrict 
ourself only to the computation of the $\ell=m=2$  ``insplunge'' 
waveform, without the matching to QNMs~\cite{Damour:2009kr}.
Note also that we neglect the non-quasi-circular corrections
advocated in Eq.~(5) of~\cite{Damour:2009kr}. 
(See also Ref.~\cite{Damour:2009ic}).  

\begin{table}[t]
\caption{\label{tab:table5} Accumulated phase difference (computed 
  from $\omega_1= 0.10799$  up to $\omega_2\equiv 2\Omega_{\rm LSO}
  = 0.13608$ ) [in radians] between $\ell=m=2$ EOB waveforms. 
  $\Delta\phi_{\rm LSO}^{\rm EOB_{5PN}}$ is the phase difference accumulated
  between the EOB$_{\rm 5PN}$ and the EOB$_{\rm testmass}$ insplunge
  waveforms, while $\Delta\phi_{\rm LSO}^{\rm EOB_{\rm 1PN}}$ is the 
  phase difference accumulated between the EOB$_{\rm 1PN}$ and the 
  EOB$_{\rm testmass}$ insplunge waveforms. 
  See text for more precise explanations.}
 \begin{center}
  \begin{ruledtabular}
  \begin{tabular}{lcc}
    $\nu$ &  $\Delta\phi_{\rm LSO}^{\rm EOB_{5PN}}$ [rad]  &  $\Delta\phi_{\rm
   LSO}^{\rm EOB_{1PN}}$ [rad]  \\
   \hline \hline 
 $10^{-2}$       & 3.2  &  0.40 \\
 $10^{-3}$       & 3.8  &  0.43 \\
 $10^{-4}$       & 4.1  &  0.44 \\
 \end{tabular}
\end{ruledtabular}
\end{center}
\end{table}
For each  value of $\nu=\{10^{-2},10^{-3},10^{-4}\}$, we compute
three insplunge $h_{22}$ resummed waveforms with increasingly
physical complexity.
The first, EOB$_{\rm testmass}$ insplunge waveform, is obtained within 
the $\O(\nu)$ approximation used so far;  i.e., we set to zero 
all the $\nu$ dependent  EOB corrections in  $H_{\rm EOB}$ 
and in the normalized flux, 
$\hat{f}_{\rm DIN}\equiv \hat{f}_{\rm DIN}(v_\varphi;\nu=0)$.
The second, EOB$_{\rm 5PN}$ insplunge waveform,  is computed from
 the full EOB dynamics, with the complete $H_{\rm EOB}$ 
and $\nu$-dependent (Newton normalized) 
flux $\hat{f}_{\rm DIN}(v_\varphi;\nu)$ replaced
in Eq.~\ref{eq:rr}. The radial potential $A(u;a_5,a_6,\nu)$ is given 
by the Pad\'e resummed form of Eq.~(2) of Ref.~\cite{Damour:2009kr}
and $a_5$ and $a_6$ are EOB flexibility parameters that take into
account 4PN and 5PN corrections in the conservative 
part of the dynamics. They have been constrained  by comparison
with numerical results~\cite{Damour:2009kr,Damour:2009sm}. 
Following~\cite{Damour:2009sm},  we use here the values 
$a_5=-22.3$ and $a_6=+252$ as ``best choice''.
The third, EOB$_{\rm 1PN}$ insplunge waveform, is obtained by keeping the same 
flux $\hat{f}_{\rm DIN}(v_\varphi;\nu)$ of 
the EOB$_{\rm 5PN}$ case, but only part of the 
EOB Hamiltonian. More precisely, we restrict the effective 
Hamiltonian $\hat{H}_{\rm eff}$ at 1PN level. 
This practically means using $A(r;0)\equiv 1-2M/r$ and 
dropping the $p_{r_*}^4/r^2$ correction term that 
enters in $\hat{H}_{\rm eff}$ at the 3PN level. 
See Eq.~(1) in~\cite{Damour:2009sm}.
We compute the relative phase difference, accumulated 
between frequencies $(\omega_1,\omega_2)$, between the
EOB$_{\rm testmass}$ waveform and the other two.
We chose $\omega_1=0.10799$, that corresponds to the initial
(test-mass) GW frequency, and $\omega_2=2\Omega_{\rm LSO}\simeq 0.13608$.
Instead of comparing the waveforms versus time, we found
it convenient  to do the following comparison versus frequency.
For each waveform, we compute the following auxiliar quantity 
\be
Q_{\omega} = \dfrac{\omega^2}{\dot{\omega}}.
\ee
This quantity measures the effective number of GW cycles spent 
around GW frequency $\omega$ (and correspondingly weighs the 
signal-to-noise ratio~\cite{Damour:2000gg}), and is a useful 
diagnostics for comparing the relative phasing accuracy of various 
waveforms~\cite{DNTrias}.
Then, the gravitational wave phase  $\phi_{(\omega_1,\omega_2)}$ 
accumulated between frequencies $(\omega_1,\omega_2)$ is 
given by
\be
\label{eq:phi}
\phi_{(\omega_1,\omega_2)}=\int_{\omega_1}^{\omega_2} Q_{\omega}d\log\omega.
\ee
We can then define the relative dephasing accumulated between two
waveforms as
\be
\Delta\phi_{(\omega_1,\omega_2)}^{{\rm EOB}_{n{\rm
      PN}}}=\int_{\omega_1}^{\omega_2}\Delta Q_{\omega}^{{\rm EOB}_{n{\rm PN}}}d\log(\omega),
\ee
where $\Delta Q_{\omega}^{{\rm EOB}_{n{\rm PN}}}\equiv Q_{\omega}^{{\rm EOB}_{n{\rm PN}}}-Q_{\omega}^{\rm EOB_{testmass}}$.
The results of this comparison are contained in Table~\ref{tab:table5}.
Note the influence of the correction due to  the conservative
part of the self force. Since this correction changes the location 
of the adiabatic $r$-LSO position~\cite{Buonanno:1998gg}, it 
entails a larger effect on the late-time portion of the 
binary dynamics and waveforms, resulting in a more 
consistent dephasing.

\section{Conclusions}
\label{sec:end}

We have presented a new calculation of the gravitational wave emission
generated through the transition from adiabatic inspiral to plunge, merger
and ringdown of a binary systems of nonspinning black holes in the extreme
mass ratio limit. We have used a Regge-Wheeler-Zerilli perturbative approach
completed by leading order EOB-based radiation reaction force. With respect
to previous work, we have improved (i) on the numerical algorithm used to
solve the Regge-Wheeler-Zerilli equations and (ii) on the analytical
definition of the improved EOB-resummed radiation reaction force.

Our main achievements are listed below.
\begin{enumerate}
\item
  We computed the complete multipolar waveform up to multipolar order $\ell=8$. 
  We focused on the relative  impact (at the level of energy 
  and angular momentum losses) of the subdominant multipoles during the 
  part of the plunge that can be considered quasiuniversal 
  (and quasigeodesic) in good approximation.
  We analyzed also the structure of the ringdown waveform 
  at the quantitative level. In particular, we measured 
  the relative amount of excitation of the fundamental QNMs with positive 
  and negative frequency. We found that, for each value of $\ell$, 
  the largest excitation of the negative modes always occurs for $m=1$ 
  and is of the order of $9\%$ of the corresponding positive mode. 

\item 
  The central numerical result of the paper is the computation of the
  gravitational recoil, or kick, imparted to the center of
  mass of the system due to the anisotropic emission of gravitational waves.
  We have discussed the influence of high modes in the multipolar expansion
  of the recoil. We showed that one has to consider $\ell\geq 4$ 
  to have a $\sim 10\%$ accuracy in the final kick.
  We found for the magnitude of the final and maximum recoil velocity
  the values $|v^{\rm end}|/\nu^2=0.0446$ and  $|v^{\rm max}|/\nu^2=0.0523$.
  The value of the final recoil shows a remarkable agreement ($<2\%$) with 
  the one extrapolated from a sample of NR simulations, 
  $|v^{\rm end}_{\rm NR}|/\nu^2\simeq 0.0439$.

\item
  The ``improved resummation'' for the radiation reaction used in this paper
  yields a better consistency agreement between mechanical angular momentum 
  losses and gravitational wave energy flux than the previously employed 
  Pad\'e resummed procedure. In particular, we found an agreement between 
  the angular momentum fluxes of the order of $0.1\%$ during the plunge 
  (well below the LSO), with a maximum disagreement of the order of $10\%$ 
  reached around the merger.
  This is a detailed piece of evidence that EOB waveforms computed via the
  resummation procedure of~\cite{Damour:2008gu} can yield accurate input 
  for LISA-oriented science.
\end{enumerate}

While writing this paper, we became aware of a similar calculation
of the final recoil by Sundararajan, Khanna and Hughes~\cite{PKH}. 
Their calculation is based on a different method to treat the 
transition from inspiral to 
plunge (see Refs.~\cite{Sundararajan:2007jg,Sundararajan:2008bw,Sundararajan:2008zm}
and references therein). In the limiting case of a nonspinning binary, 
their results for the final and maximum kick are fully consistent with ours.

\acknowledgements

We thank Thibault Damour for fruitful discussions, 
inputs and a careful reading of the manuscript.
We also acknowledge useful correspondence with 
Scott Hughes, Gaurav Khanna and Pranesh Sundararajan, 
who made us kindly aware of their results before publication.
We are grateful to Bernd Br\"ugmann, Mark Hannam, Sascha Husa, 
Jos\'e~A.~Gonz\'alez, and Ulrich Sperhake for giving us access to their NR data.
Computations were performed on the INFN Beowulf clusters 
{\tt Albert} at the University of Parma and the {\tt Merlin} 
cluster at IHES. We also thank Roberto De Pietri, Fran\c{c}ois Bachelier, and
Karim Ben Abdallah for technical assistance 
and E.~Berti for discussions.
SB is supported by DFG Grant SFB/Transregio~7 ``Gravitational Wave Astronomy''. 
SB thank IHES for hospitality and support 
during the development of this work.

\appendix

\section{Numerical framework, tests and comparison with the literature}
\label{sec:numerical}

The numerical procedure adopted is similar to the one of Ref.~\cite{Nagar:2006xv},
but it has been improved on several aspects. In particular, the original code 
has been fully rewritten and optimized and a new finite-differencing 
algorithm to solve the Regge-Wheeler-Zerilli equation has been implemented. 

The Regge-Wheeler-Zerilli equations, Eqs.~\eqref{eq:rwz}, are solved as 
a first-order-in-time second-order-in-space system adopting the method 
of lines. Time advancing is done by means of a 4th order Runge-Kutta 
algorithm, while centered 4th order finite differences are used to 
approximate the space derivative. Standard Sommerfeld-maximally dissipative 
boundary conditions are adopted and implemented as described 
in~\cite{Calabrese:2005fp}
\begin{figure}[t]
\begin{center}
  \includegraphics[width=0.49\textwidth]{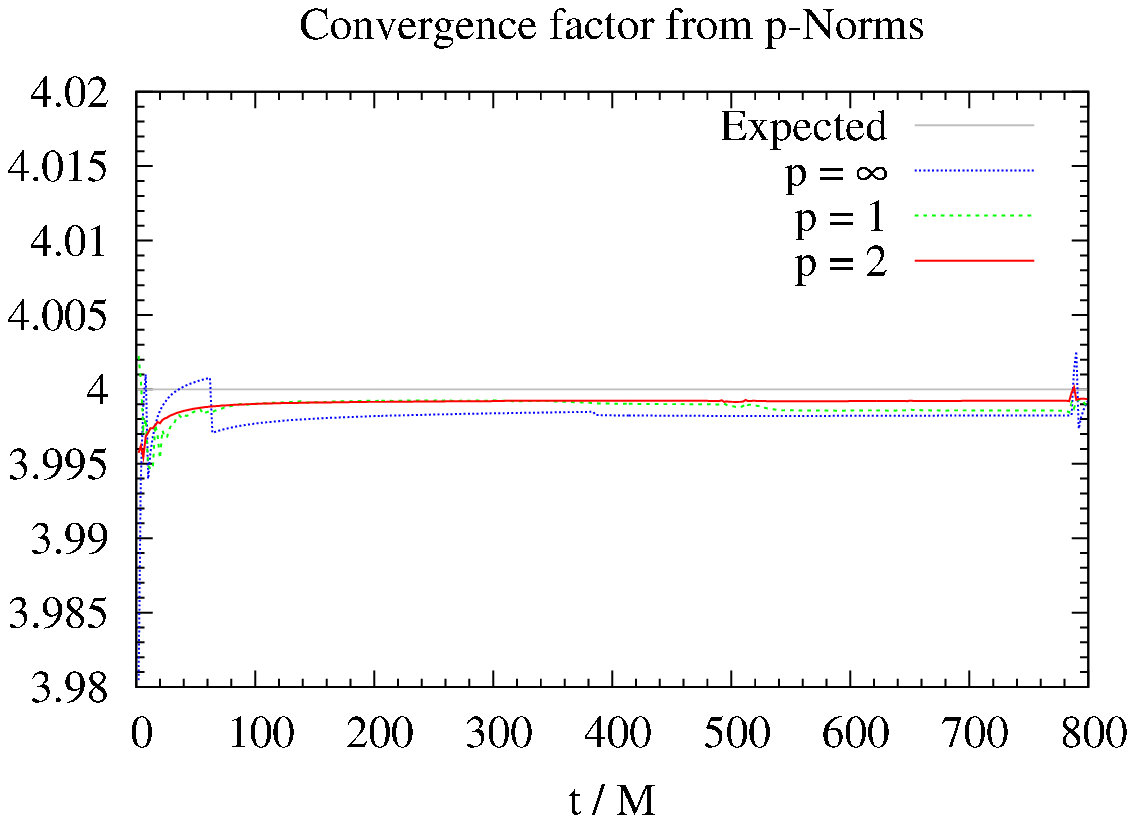}
  \caption{\label{fig:convnorm} Self-convergence factor computed from 
    $p$-norms of $(\ell=8,m=8)$ inspiral plunge waveforms. 
    Resolutions used are: $\Delta r_*=\{0.05,\, 0.025,\, 0.0125\}$. 
    The particle is initially at $R(0)=6.5M$ and $\nu=0.01$.}
\end{center} 
\end{figure}

We solve the equations given in Sec.~\ref{sec:dynamics} for the particle 
dynamics using a standard 4th order Runge-Kutta algorithm with adaptive 
stepsize. Then we insert the resulting position and momenta in
the  source terms $S_{\ell m}^{(\rm e/o)}$ using a Gaussian-function 
representation of $\delta(r_*-R_*(t))$ (see below).
The distributional $\delta$-function that 
appears in the source terms is approximated by a smooth function 
$\delta_\sigma(r_*)$. We use
\begin{equation}
\label{eq:delta}
\delta(r_*-R_*(t))\rightarrow \delta_\sigma(r_*)=\frac{1}{\sigma\sqrt{2\pi}}\exp \left[{-\frac{(r_*-R_*(t))^2}{2\sigma^2}}\right] \ ,
\end{equation}
with $\sigma\geq\Delta r_*$. In practice $\sigma\simeq\Delta r_*$ works well thanks to
the effective averaging entailed by the fact that $R_*(t)$ is not restricted 
to the $r_*$ grid, but varies nearly continuously on the $r_*$ axis. 
In Ref.~\cite{Nagar:2006xv} it was already pointed out that, if $\sigma$ is
sufficiently small and the resolution sufficiently high (so that the Gaussian
function is resolved by a sufficiently high number of points) this
technique is competitive with other approaches that prefer a mathematically 
more rigorous treatment of the $\delta$-function~\cite{Lousto:1997wf,Martel:2001yf,Sopuerta:2005gz}
(see in this respect Table~1 in Ref.~\cite{Nagar:2006xv}). 
Since in this paper we use a different numerical method to solve
Eqs.~\eqref{eq:rwz}, we have performed exnovo all the accuracy tests
for circular orbits and radial plunge that were formerly discussed
in~\cite{Nagar:2006xv}.

Self-convergence tests showed the correct convergence rate both in norm, 
see Fig.~\ref{fig:convnorm} for an example, and pointwise, both with and without 
the particle source.
In the latter case however results were not satisfactory if the Gaussian in 
the source was not enough resolved. We found the correct convergence rate
using, for the lowest resolution, a Gaussian width of $\sigma\geq3\Delta r_*$, 
optimal results were obtained with $\sigma\sim10\Delta r_*$, while 
smaller values gives experimental rate around 2nd ($\sigma=\Delta r_*$) and 
3rd order  ($\sigma=2\Delta r_*$). Together with the physical requirement 
$\sigma\ll M$ and the necessity of extracting waveforms at large radii,
this fact poses some limits on the resolution to be used and on the minimal 
computational time necessary for the simulations.
As expected, no spurious oscillations were found in the Regge-Wheeler-Zerilli 
solution in the region across the (smoothed) delta function.

\begin{figure}[t]
\begin{center}
  \includegraphics[width=0.49\textwidth]{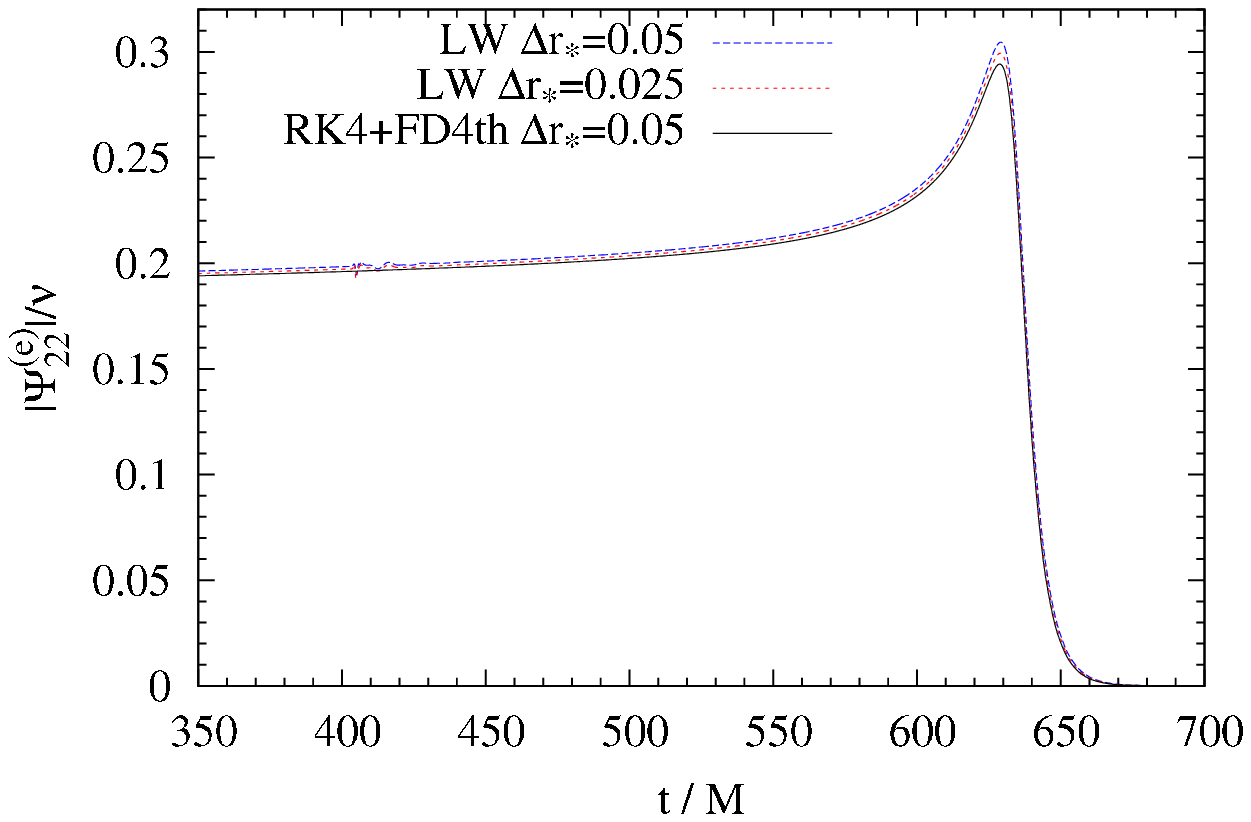}
  \caption{\label{fig:oldnewcode} Comparison between the old and the new code.
    The plot shows the amplitude of the $(2,2)$ mode of the
    Regge-Wheeler-Zerilli as computed with the old code 
    (at resolutions $\Delta r_*=0.5$ and $0.25$) based on the Lax-Wendroff 
    scheme and with the new code (at resolution $\Delta r_*=0.5$).
    Other parameters of the runs are $\mu=0.001$, $R(0)=6.5$ and $c_{\rm cfl}=0.9$.}
\end{center} 
\end{figure}

A direct comparison with the old code, Fig.~\ref{fig:oldnewcode}, 
shows clearly that the numerical improvements lead to quantitative 
better results. The amplitude of $\Psi_{2\,2}$ is computed for 
two different resolution $\Delta r_*=0.5$ and $0.25$ using the old code  
based on the Lax-Wendroff scheme and with $\Delta r_*=0.5$ with the new scheme. 

The differences are due only to the numerical scheme employed for the
wave equation, since also the new radiation reaction has been used 
in the old code. As expected, the new numerical scheme shows a 
faster convergence and strongly suppresses the spurious oscillations 
coming from the boundaries; note in this respect the small 
``bumps'' at $t\sim400$ that are present only in the data 
computed with the old code.

To validate the physical results of the code at a more quantitative 
level we performed ``standard'' comparisons with the literature
considering circular orbits and radial plunge, following the line
of~\cite{Nagar:2006xv}.
\begin{table*}[t]
  \caption{\label{tab:ciro:E} Gravitational wave (multipolar) energy flux $\dot{E}_{\ell m}/\mu^2$ 
    of a particle on a circular orbit of radius
    $r_0=7.9456M$. Comparison between our results and those present in the literature.
    Our waveforms are extracted at $r_*=1000 M$.}
  \begin{center}
  \begin{ruledtabular}
    \begin{tabular}{cccccccc}
      $\ell$ $m$ & This work 
      & Ref.~\cite{Sopuerta:2005gz}%
      & Diff.[\%] %
      & Ref.~\cite{Martel:2003jj}%
      & Diff.[\%]%
      & Ref.~\cite{Poisson}%
      & Diff.[\%]\\
      \hline \hline 
      2 1& 8.1733$\times10^{-7}$& 8.1662$\times10^{-7}$& 0.086& 8.1623$\times10^{-7}$& 0.134& 8.1633$\times10^{-7}$& 0.122\\
      2 2& 1.7069$\times10^{-4}$& 1.7064$\times10^{-4}$& 0.029& 1.7051$\times10^{-4}$& 0.105& 1.7063$\times10^{-4}$& 0.035\\
      3 1& 2.1785$\times10^{-9}$& 2.1732$\times10^{-9}$& 0.242& 2.1741$\times10^{-9}$& 0.200& 2.1731$\times10^{-9}$& 0.246\\
      3 2& 2.5218$\times10^{-7}$& 2.5204$\times10^{-7}$& 0.057& 2.5164$\times10^{-7}$& 0.217& 2.5199$\times10^{-7}$& 0.077\\
      3 3& 2.5483$\times10^{-5}$& 2.5475$\times10^{-5}$& 0.031& 2.5432$\times10^{-5}$& 0.201& 2.5471$\times10^{-5}$& 0.047\\
      4 1& 8.3699$\times10^{-13}$& 8.4055$\times10^{-13}$& 0.423& 8.3507$\times10^{-13}$& 0.230& 8.3956$\times10^{-13}$& 0.306\\
      4 2& 2.5125$\times10^{-9}$& 2.5099$\times10^{-9}$& 0.103& 2.4986$\times10^{-9}$& 0.556& 2.5091$\times10^{-9}$& 0.135\\
      4 3& 5.7792$\times10^{-8}$& 5.7765$\times10^{-8}$& 0.046& 5.7464$\times10^{-8}$& 0.571& 5.7751$\times10^{-8}$& 0.071\\
      4 4& 4.7283$\times10^{-6}$& 4.7270$\times10^{-6}$& 0.027& 4.7080$\times10^{-6}$& 0.430& 4.7256$\times10^{-6}$& 0.056\\
      5 1& 1.2904$\times10^{-15}$& 1.2607$\times10^{-15}$& 2.357& 1.2544$\times10^{-15}$& 2.871& 1.2594$\times10^{-15}$& 2.462\\
      5 2& 2.7874$\times10^{-12}$& 2.7909$\times10^{-12}$& 0.126& 2.7587$\times10^{-12}$& 1.040& 2.7896$\times10^{-12}$& 0.080\\
      5 3& 1.0946$\times10^{-9}$& 1.0936$\times10^{-9}$& 0.095& 1.0830$\times10^{-9}$& 1.074& 1.0933$\times10^{-9}$& 0.122\\
      5 4& 1.2334$\times10^{-8}$& 1.2329$\times10^{-8}$& 0.038& 1.2193$\times10^{-8}$& 1.154& 1.2324$\times10^{-8}$& 0.078\\
      5 5& 9.4630$\times10^{-7}$& 9.4616$\times10^{-7}$& 0.014& 9.3835$\times10^{-7}$& 0.847& 9.4563$\times10^{-7}$& 0.071\\
    \end{tabular}
  \end{ruledtabular}
 \end{center}
\end{table*}

The energy and angular  momentum fluxes computed from the 
waveforms generated by a particle on a circular orbit of radius
$r_0=7.9456M$ are displayed in Table~\ref{tab:ciro:E} and 
Table~\ref{tab:ciro:J}. The numbers are compared with those present 
in the literature, showing very good agreement. 
The fractional differences are always well below the $1\%$
except for the multipole $(5,1)$ ($2\%$ in the energy flux) 
whose absolute value is the smallest. Notice that, differently 
from Ref.~\cite{Nagar:2006xv}, the accuracy is maintained also 
for high multipoles. We also computed multipoles for $\ell\geq 6$,
although we did not report them here since corresponding data
 to compare with are not present in the literature.
\begin{figure}[t]
\begin{center}
  \includegraphics[width=0.49\textwidth]{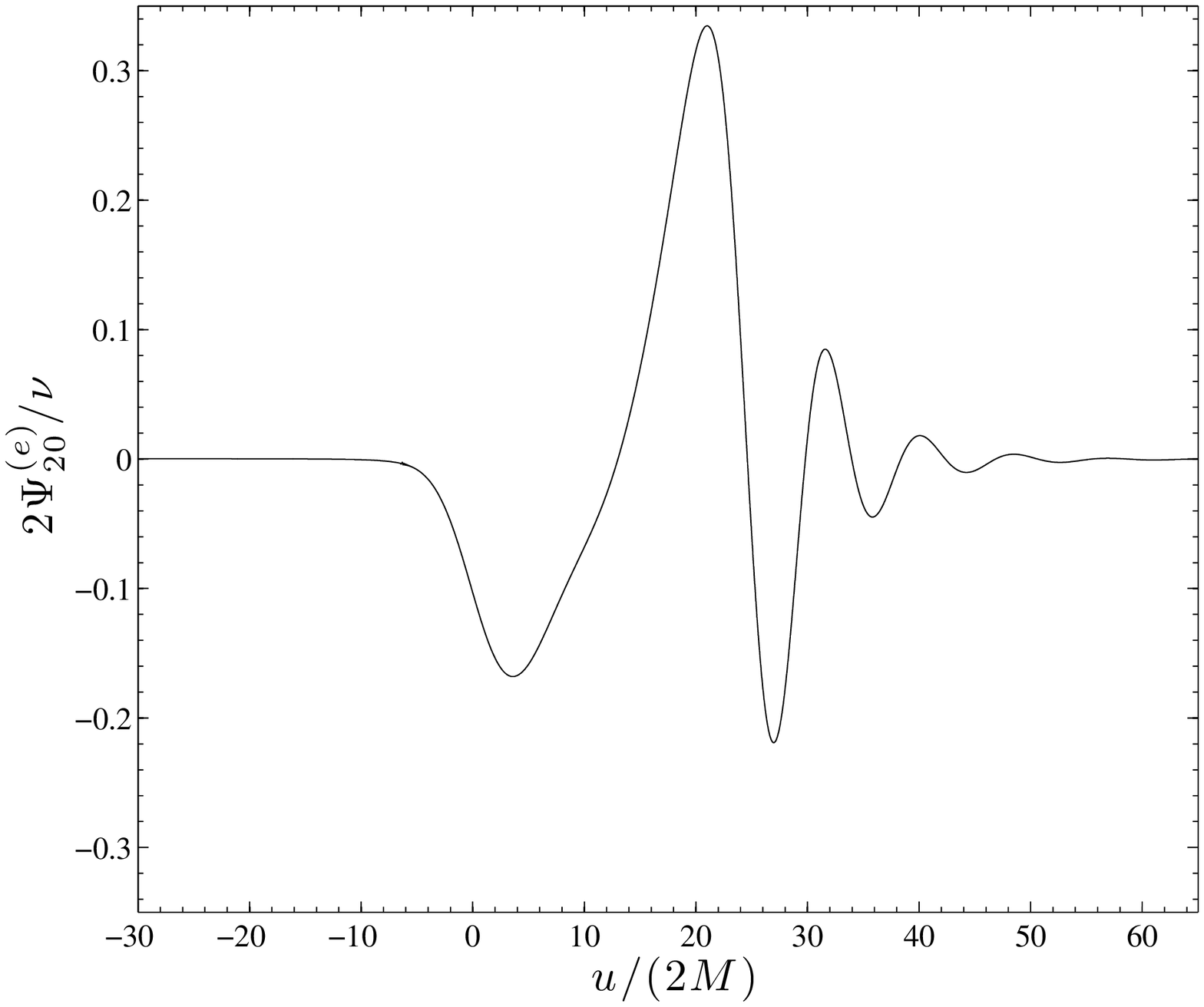}
  \caption{\label{fig:fig12}Waveform emitted by a particle plunging radially
  into the black hole (along the $z$-axis) from $r=10M$. The waveform is
  extracted at $r_*=1000M$.}
\end{center} 
\end{figure}

For what concerns the radial infall, Fig.~\ref{fig:fig12} displays the
$\ell=2$, $m=0$ waveform generated by a particle plunging into the black
hole radially along the $z$-axis. The particle has zero initial velocity
and starts at $r=10M$. We specify conformally flat initial data according
to the procedure described at the end of Sec.~4 of~\cite{Nagar:2006xv}.
Note that $\Psi^{(\rm e)}_{20}$ has been multiplied by a factor 2 to 
facilitate the (very satisfactory) comparison with the top-right panel of 
Fig.~4 in~\cite{Martel:2001yf} and top-left 
panel of Fig.~6 in~\cite{Lousto:1997wf}.

\begin{table*}[t]
  \caption{\label{tab:ciro:J} Gravitational wave (multipolar) angolar momentum 
    flux $\dot{J}_{\ell m} / \mu^2$ of a particle on a circular orbit of radius
    $r_0=7.9456M$. Comparison between our results and those present in the literature.
    Our waveforms are extracted at $r_*=1000 M$.}
  \begin{center}
    \begin{ruledtabular}
      \begin{tabular}{ccccccccc}
        $\ell$ $m$ & This work 
        & Ref.~\cite{Sopuerta:2005gz} %
        & Diff.[\%] 
        & Ref.~\cite{Martel:2003jj} %
        & Diff.[\%] 
        & Ref.~\cite{Poisson} %
        & Diff.[\%]\\
        \hline \hline 
        2 1& 1.8305$\times10^{-5}$& 1.8289$\times10^{-5}$& 0.090& 1.8270$\times10^{-5}$& 0.194& 1.8283$\times10^{-5}$& 0.122\\
        2 2& 3.8229$\times10^{-3}$& 3.8219$\times10^{-3}$& 0.027& 3.8164$\times10^{-3}$& 0.171& 3.8215$\times10^{-3}$& 0.037\\
        3 1& 4.8790$\times10^{-8}$& 4.8675$\times10^{-8}$& 0.237& 4.8684$\times10^{-8}$& 0.219& 4.8670$\times10^{-8}$& 0.247\\
        3 2& 5.6481$\times10^{-6}$& 5.6450$\times10^{-6}$& 0.055& 5.6262$\times10^{-6}$& 0.389& 5.6439$\times10^{-6}$& 0.075\\
        3 3& 5.7074$\times10^{-4}$& 5.7057$\times10^{-4}$& 0.030& 5.6878$\times10^{-4}$& 0.345& 5.7048$\times10^{-4}$& 0.046\\
        4 1& 1.8518$\times10^{-11}$& 1.8825$\times10^{-11}$& 1.633& 1.8692$\times10^{-11}$& 0.933& 1.8803$\times10^{-11}$& 1.517\\
        4 2& 5.6272$\times10^{-8}$& 5.6215$\times10^{-8}$& 0.102& 5.5926$\times10^{-8}$& 0.619& 5.6195$\times10^{-8}$& 0.138\\
        4 3& 1.2944$\times10^{-6}$& 1.2937$\times10^{-6}$& 0.052& 1.2933$\times10^{-6}$& 0.083& 1.2934$\times10^{-6}$& 0.075\\
        4 4& 1.0590$\times10^{-4}$& 1.0586$\times10^{-4}$& 0.037& 1.0518$\times10^{-4}$& 0.684& 1.0584$\times10^{-4}$& 0.056\\
        5 1& 2.8558$\times10^{-14}$& 2.8237$\times10^{-14}$& 1.138& 2.8090$\times10^{-14}$& 1.667& 2.8206$\times10^{-14}$& 1.249\\
        5 2& 6.2386$\times10^{-11}$& 6.2509$\times10^{-11}$& 0.197& 6.1679$\times10^{-11}$& 1.146& 6.2479$\times10^{-11}$& 0.149\\
        5 3& 2.4517$\times10^{-8}$& 2.4494$\times10^{-8}$& 0.093& 2.4227$\times10^{-8}$& 1.196& 2.4486$\times10^{-8}$& 0.125\\
        5 4& 2.7625$\times10^{-7}$& 2.7613$\times10^{-7}$& 0.042& 2.7114$\times10^{-7}$& 1.883& 2.7603$\times10^{-7}$& 0.078\\
        5 5& 2.1194$\times10^{-5}$& 2.1190$\times10^{-5}$& 0.020& 2.0933$\times10^{-5}$& 1.248& 2.1179$\times10^{-5}$& 0.072\\
      \end{tabular}
    \end{ruledtabular}
  \end{center}
\end{table*}

An additional test involved the dependence of the results on the
Gaussian ampitude $\sigma$. Focusing on $\mu=0.01$ and inspiral plunge
simulations with $R(0)=6.5$ M, we experimented with the values
$\sigma=\{0.05,0.1,0.5,1\}$ M using typical resolutions. 
The extreme value $\sigma=M$ (see discussion above) gave
reliable waveforms, while spurious modulations due to the extended 
source were clearly evident for $\sigma>M$.
Relative differences between $(2,2)$-waveforms, taking $\sigma=0.05$ M as
reference value, were of the order of
$\{10^{-2},\,10^{-3},\,10^{-4}\}$ in the amplitudes respectively for $\sigma=\{0.01,0.5,1\}$ M,
and 1 order of magnitude less for the phase. 
For the $(8,8)$ multipole (worst case) they were $\{10^{-1},\,10^{-2},\,10^{-3}\}$
always for $\sigma=\{0.01,0.5,1\}$ M.
The relative differences on the energy flux multipoles were of the
order of $\{0.0002,\,0.007,\,0.03\}$ in the $(2,2)$ case and  
$\{0.005,\,0.2,\,0.9\}$ in the $(8,8)$ case.
While $\sigma\geq0.5$ M does not give satisfactory results, 
differences between $\sigma=0.05$ and $\sigma=0.1$ are quite small,
giving the same results up to $0.5\%$ for the energy contribute.
In the simulations of the paper we used $\sigma=0.04M$. 

We also tested the use of the alternative source term (mathematically
equivalent for a distributional source) given by Eq~(23) 
of~\cite{Nagar:2006xv}. Consistently with the analysis of 
Ref.~\cite{Nagar:2006xv} (see their Fig.~5), for insplunge waveforms
 we found a  relative difference smaller than $10^{-6}$ both in 
amplitude and phase in the $\ell=m=2$ mode.

We finally mention that in the case of long simulations 
(i.e., $\nu=10^{-3}$, $r_0=7M$) our boundary conditions 
are not fully satisfactory. 
In fact, in this case the waveforms might be (slightly) contaminated 
by small reflections from boundaries, especially for high multipoles.
A solution to this problem is discussed in~\cite{Lau:2005ti} 
(and references therein), in which, basically, the free-data in 
the Sommerfeld condition (in our case they are set to zero) are 
specified as an integral convolution between a time-domain boundary kernel 
and the solution. The method provides an exact radiative outer boundary 
condition for the wave equations.
We mention that an alternative approach is represented by solving the
Regge-Wheeler-Zerilli equation on matched hyperboloidal foliations
as described in~\cite{Zenginoglu:2009ey} (and references therein).
The appealing feature is the fact that it has the double advantage to
not need boundary conditions (no incoming modes) and 
to allow extraction exactly at null infinity.

\section{Partial losses during the plunge phase}
\label{sec:losses}

Let us list here some useful numerical information related to the
discussion of Sec.~\ref{sec:qgeoplunge}; i.e, the energy and 
angular momentum emitted during the quasiuniversal part of the late-plunge,
merger and ringdown. The pure numbers, up to multuipolar order $\ell=8$, 
are given in Table~\ref{tab_app1}. The relative percentage (with respect
to the ``total'' energy  $E_{\rm TOT}$ and angular momentum $J_{\rm TOT}$)
are given in the following Table~\ref{tab_app2}. Note here that by
total we indicate the sum over multipoles $(\ell,m)$, with 
$2\leq \ell\leq 8$ and $|m|\leq \ell$. 
\begin{table*}[t]
\caption{\label{tab_app1}Multipolar contributions to total energy
and angular momentum emitted during the quasiuniversal part of 
the plunge phase, the merger and the ringdown.}
 \begin{center}
  \begin{ruledtabular}
  \begin{tabular}{cc|cc|cc}
    \multicolumn{2}{c|}{Multipole} & 
    \multicolumn{2}{c|}{$\mu=0.001$} &
    \multicolumn{2}{c}{$\mu=0.0001$} \\
    $\ell$ & $m$ & $ME_{\ell m}/\mu^2$  & $J_{\ell m}/\mu^2$ 
    & $ME_{\ell m}/\mu^2$  & $J_{\ell m}/\mu^2$ \\
   \hline \hline 
2 & 0  & 9.739$\times 10^{-4}$  &  0  & 9.739$\times 10^{-4}$ & 0 \\
  & 1 & 2.023 $\times 10^{-2}$ & 0.787 $\times 10^{-1}$ & 2.035 $\times 10^{-2}$ & 0.793 $\times 10^{-1}$ \\ 
  & 2 & 2.733 $\times 10^{-1}$ & 2.150                 & 2.769 $\times 10^{-1}$ & 2.179  \\ 
3 & 0 & 3.320 $\times 10^{-5}$ &   0                   & 3.330$\times 10^{-5}$   & 0                     \\
  & 1 & 0.540 $\times 10^{-3}$ & 1.196 $\times 10^{-3}$ & 0.543 $\times 10^{-3}$ & 1.201 $\times 10^{-3}$ \\ 
  & 2 & 0.782 $\times 10^{-2}$ & 0.370 $\times 10^{-1}$ & 0.788 $\times 10^{-2}$ & 0.373 $\times 10^{-1}$ \\ 
  & 3 & 0.929 $\times 10^{-1}$ & 0.678  & 0.941 $\times 10^{-1}$ & 0.687  \\ 
4 & 0 & 1.518 $\times 10^{-6}$ &                0      & 1.523$\times 10^{-6}$   & 0                      \\
  & 1 & 2.226 $\times 10^{-5}$ & 0.333 $\times 10^{-4}$ & 2.234 $\times 10^{-5}$ & 0.334 $\times 10^{-4}$ \\ 
  & 2 & 3.023 $\times 10^{-4}$ & 0.998 $\times 10^{-3}$ & 3.039 $\times 10^{-4}$ & 1.003 $\times 10^{-3}$ \\ 
  & 3 & 0.330 $\times 10^{-2}$ & 1.657 $\times 10^{-2}$ & 0.332 $\times 10^{-2}$ & 1.674 $\times 10^{-2}$ \\ 
  & 4 & 0.377 $\times 10^{-1}$ & 2.627 $\times 10^{-1}$ & 0.382 $\times 10^{-1}$ & 2.663 $\times 10^{-1}$ \\ 
5 & 0 & 7.932 $\times 10^{-8}$ &                 0     & 7.957 $\times 10^{-8}$  & 0                     \\
  & 1 & 1.042 $\times 10^{-6}$ & 1.205 $\times 10^{-6}$ & 1.046 $\times 10^{-6}$ & 1.209 $\times 10^{-6}$ \\ 
  & 2 & 1.341 $\times 10^{-5}$ & 0.323 $\times 10^{-4}$ & 1.347 $\times 10^{-5}$ & 0.325 $\times 10^{-4}$ \\ 
  & 3 & 1.659 $\times 10^{-4}$ & 0.642 $\times 10^{-3}$ & 1.670 $\times 10^{-4}$ & 0.647 $\times 10^{-3}$ \\ 
  & 4 & 1.478 $\times 10^{-3}$ & 0.762 $\times 10^{-2}$ & 1.493 $\times 10^{-3}$ & 0.770 $\times 10^{-2}$ \\ 
  & 5 & 1.683 $\times 10^{-2}$ & 1.129 $\times 10^{-1}$ & 1.705 $\times 10^{-2}$ & 1.145 $\times 10^{-1}$ \\ 
6 & 0  & 4.182$\times 10^{-9}$ &  0  & 4.197$\times 10^{-9}$ & 0 \\
  & 1 & 0.551 $\times 10^{-7}$ & 0.516 $\times 10^{-7}$ & 0.552 $\times 10^{-7}$ & 0.518 $\times 10^{-7}$                  \\ 
  & 2 & 0.673 $\times 10^{-6}$ & 1.312 $\times 10^{-6}$ & 0.676 $\times 10^{-6}$ & 1.317 $\times 10^{-6}$ \\ 
  & 3 & 0.782 $\times 10^{-5}$ & 2.340 $\times 10^{-5}$ & 0.786 $\times 10^{-5}$ & 2.352 $\times 10^{-5}$ \\ 
  & 4 & 0.904 $\times 10^{-4}$ & 0.379 $\times 10^{-3}$ & 0.910 $\times 10^{-4}$ & 0.382 $\times 10^{-3}$ \\ 
  & 5 & 0.694 $\times 10^{-3}$ & 0.361 $\times 10^{-2}$ & 0.701 $\times 10^{-3}$ & 0.365 $\times 10^{-2}$ \\ 
  & 6 & 0.799 $\times 10^{-2}$ & 0.518 $\times 10^{-1}$ & 0.809 $\times 10^{-2}$ & 0.526 $\times 10^{-1}$ \\ 
7 & 0  & 2.416$\times 10^{-10}$  &  0  & 2.424 $\time 10^{-10}$ & 0 \\
  & 1 & 2.948 $\times 10^{-9}$ & 2.337 $\times 10^{-9}$ & 2.959 $\times 10^{-9}$ & 2.345 $\times 10^{-9}$ \\ 
  & 2 & 0.360 $\times 10^{-7}$ & 0.586 $\times 10^{-7}$ & 0.362 $\times 10^{-7}$ & 0.588 $\times 10^{-7}$ \\ 
  & 3 & 0.423 $\times 10^{-6}$ & 1.058 $\times 10^{-6}$ & 0.425 $\times 10^{-6}$ & 1.063 $\times 10^{-6}$ \\ 
  & 4 & 0.448 $\times 10^{-5}$ & 1.516 $\times 10^{-5}$ & 0.451 $\times 10^{-5}$ & 1.526 $\times 10^{-5}$ \\ 
  & 5 & 0.492 $\times 10^{-4}$ & 2.168 $\times 10^{-4}$ & 0.496 $\times 10^{-4}$ & 2.186 $\times 10^{-4}$ \\ 
  & 6 & 0.337 $\times 10^{-3}$ & 1.760 $\times 10^{-3}$ & 0.341 $\times 10^{-3}$ & 1.781 $\times 10^{-3}$ \\ 
  & 7 & 0.396 $\times 10^{-2}$ & 2.497 $\times 10^{-2}$ & 0.401 $\times 10^{-2}$ & 2.533 $\times 10^{-2}$ \\ 
8 & 0  & 1.359 $\times 10^{-11}$  &  0  & 1.364 $\times 10^{-11}$ & 0 \\
  & 1 & 1.699 $\times 10^{-10}$ & 1.163 $\times 10^{-10}$ &  1.704$\times 10^{-10}$ &  1.167$\times 10^{-10}$  \\ 
  & 2 & 1.986 $\times 10^{-9}$ & 2.792 $\times 10^{-9}$ & 1.994 $\times 10^{-9}$ & 2.802 $\times 10^{-9}$ \\ 
  & 3 & 2.303 $\times 10^{-8}$ & 0.491 $\times 10^{-7}$ & 2.313 $\times 10^{-8}$ & 0.493 $\times 10^{-7}$ \\ 
  & 4 & 2.604 $\times 10^{-7}$ & 0.756 $\times 10^{-6}$ & 2.618 $\times 10^{-7}$ & 0.760 $\times 10^{-6}$ \\ 
  & 5 & 2.539 $\times 10^{-6}$ & 0.931 $\times 10^{-5}$ & 2.558 $\times 10^{-6}$ & 0.938 $\times 10^{-5}$ \\ 
  & 6 & 2.696 $\times 10^{-5}$ & 1.223 $\times 10^{-4}$ & 2.721 $\times 10^{-5}$ & 1.235 $\times 10^{-4}$ \\ 
  & 7 & 1.680 $\times 10^{-4}$ & 0.878 $\times 10^{-3}$ & 1.701 $\times 10^{-4}$ & 0.890 $\times 10^{-3}$ \\ 
  & 8 & 2.026 $\times 10^{-3}$ & 1.248 $\times 10^{-2}$ & 2.054 $\times 10^{-3}$ & 1.266 $\times 10^{-2}$ \\ 
\end{tabular}
\end{ruledtabular}
\end{center}
\end{table*}
\begin{table*}[t]
\caption{\label{tab_app2}Relative contribution of each multipole
to the total energy and angular momentum emitted during the
quasiuniversal part of the plunge phase, the merger 
and the ringdown.}
 \begin{center}
  \begin{ruledtabular}
  \begin{tabular}{cc|cc|cc}
    \multicolumn{2}{c|}{Multipole} & 
    \multicolumn{2}{c|}{$\mu=0.001$} &
    \multicolumn{2}{c}{$\mu=0.0001$} \\
    $\ell$ & $m$ & $E_{\ell m}/E_{\rm TOT}[\%]$  & $J_{\ell m}/J_{\rm TOT} [\%]$& 
    $E_{\ell m}/E_{\rm TOT}$ $[\%]$  &
    $J_{\ell m}/J_{\rm TOT}$ $[\%]$ \\
   \hline \hline 
2 & 0  & 0.2068  &  0            & 0.2050   & 0        \\
  & 1  & 4.295   & 2.2868        &  4.268   & 2.2730   \\ 
  & 2  & 58.03   & 62.45         & 58.07    & 62.46    \\ 
3 & 0  & 7.0504 $\times 10^{-3}$  &  0      & 6.9844 $\times 10^{-3}$ & 0 \\
  & 1  & 1.1475 $\times 10^{-1}$  & 0.3475 $\times 10^{-1}$ & 1.1378 $\times 10^{-1}$ & 0.3443 $\times 10^{-1}$ \\ 
  & 2  & 1.6609                  & 1.0735  & 1.6530 & 1.0689 \\ 
  & 3  & 19.724                  & 19.695  & 19.726 & 19.698  \\ 
4 & 0  & 3.2237$\times 10^{-4}$   &  0  & 3.1947$\times 10^{-4}$ & 0 \\
  & 1 & 0.4727 $\times 10^{-2}$   & 0.9683 $\times 10^{-3}$ & 0.4685 $\times 10^{-2}$ & 0.9586 $\times 10^{-3}$ \\ 
  & 2 & 0.6419 $\times 10^{-1}$   & 0.2898 $\times 10^{-1}$ & 0.6372 $\times 10^{-1}$ & 0.2875 $\times 10^{-1}$ \\ 
  & 3 & 0.6997                   & 0.4814                 & 0.6972                  & 0.4799                  \\ 
  & 4 & 8.006                    & 7.631                  & 8.007                   & 7.633  \\ 
5 & 0 & 1.6843 $\times 10^{-5}$  &  0  & 1.6686$\times 10^{-5}$  & 0 \\
  & 1 & 2.2133 $\times 10^{-4}$  & 0.3501 $\times 10^{-4}$ & 2.1936 $\times 10^{-4}$ & 0.3466 $\times 10^{-4}$ \\ 
  & 2 & 2.8483 $\times 10^{-3}$  & 0.9389 $\times 10^{-3}$ & 2.8252 $\times 10^{-3}$ & 0.9305 $\times 10^{-3}$ \\ 
  & 3 & 0.3523 $\times 10^{-1}$  & 0.1865 $\times 10^{-1}$ & 0.3501 $\times 10^{-1}$ & 0.1853 $\times 10^{-1}$ \\ 
  & 4 & 0.3139                  & 0.2213                 & 0.3130                & 0.2208                  \\ 
  & 5 & 3.574                   & 3.280                 & 3.575                 & 3.281                  \\ 
6 & 0  & 8.8792$\times 10^{-7}$   &  0  & 8.8013$\times 10^{-7}$ & 0 \\
  & 1 & 1.1691 $\times 10^{-5}$  & 1.4994 $\times 10^{-6}$ & 1.1585 $\times 10^{-5}$ & 1.148428 $\times 10^{-6}$ \\ 
  & 2 & 1.4293 $\times 10^{-4}$  & 0.3810 $\times 10^{-4}$ & 1.4172 $\times 10^{-4}$ & 0.3774 $\times 10^{-4}$ \\ 
  & 3 & 1.6611 $\times 10^{-3}$  & 0.6796 $\times 10^{-3}$ & 1.6492 $\times 10^{-3}$ & 0.6742 $\times 10^{-3}$ \\ 
  & 4 & 1.9185 $\times 10^{-2}$  & 1.1019 $\times 10^{-2}$ & 1.9085 $\times 10^{-2}$ & 1.0957 $\times 10^{-2}$ \\ 
  & 5 & 1.4729 $\times 10^{-1}$  & 1.0489 $\times 10^{-1}$ & 1.4703 $\times 10^{-1}$ & 1.0470 $\times 10^{-1}$ \\ 
  & 6 & 1.6963                  & 1.5060                  & 1.6970                 & 1.5069  \\ 
7 & 0  & 5.1297$\times 10^{-8}$ &  0  & 5.0836$\times 10^{-8}$ & 0 \\
  & 1 & 0.6260 $\times 10^{-6}$ & 0.6787 $\times 10^{-7}$ & 0.6205 $\times 10^{-6}$ & 0.6720 $\times 10^{-7}$ \\ 
  & 2 & 0.7651 $\times 10^{-5}$ & 1.7025 $\times 10^{-6}$ & 0.7585 $\times 10^{-5}$ & 1.6861 $\times 10^{-6}$ \\ 
  & 3 & 0.8986 $\times 10^{-4}$ & 3.0728 $\times 10^{-5}$ & 0.8916 $\times 10^{-4}$ & 3.0458 $\times 10^{-5}$ \\ 
  & 4 & 0.9508 $\times 10^{-3}$ & 0.4404 $\times 10^{-3}$ & 0.9450 $\times 10^{-3}$ & 0.4373 $\times 10^{-3}$ \\ 
  & 5 & 1.0456 $\times 10^{-2}$ & 0.6296 $\times 10^{-2}$ & 1.0411 $\times 10^{-2}$ & 0.6266 $\times 10^{-2}$ \\ 
  & 6 & 0.7152 $\times 10^{-1}$ & 0.5111 $\times 10^{-1}$ & 0.7145 $\times 10^{-1}$ & 0.5106 $\times 10^{-1}$ \\ 
  & 7 & 0.8405                  & 0.7254 & 0.8412  & 0.7260  \\ 
8 & 0  & 2.8851$\times 10^{-9}$  &  0  & 2.8611$\times 10^{-9}$ & 0 \\
  & 1 & 0.3608 $\times 10^{-7}$  & 0.3378 $\times 10^{-8}$ & 0.3574 $\times 10^{-7}$ & 0.3344 $\times 10^{-8}$ \\ 
  & 2 & 0.4217 $\times 10^{-6}$  & 0.8109 $\times 10^{-7}$ & 0.4180 $\times 10^{-6}$ & 0.8031 $\times 10^{-7}$ \\ 
  & 3 & 0.4889 $\times 10^{-5}$  & 1.4261 $\times 10^{-6}$ & 0.4850 $\times 10^{-5}$ & 1.4132 $\times 10^{-6}$ \\ 
  & 4 & 0.5529 $\times 10^{-4}$  & 2.1963 $\times 10^{-5}$ & 0.5490 $\times 10^{-4}$ & 2.1788 $\times 10^{-5}$ \\ 
  & 5 & 0.5391 $\times 10^{-3}$  & 2.7038 $\times 10^{-4}$ & 0.5364 $\times 10^{-3}$ & 2.6880 $\times 10^{-4}$ \\ 
  & 6 & 0.5725 $\times 10^{-2}$  & 0.3553 $\times 10^{-2}$ & 0.5706 $\times 10^{-2}$ & 0.3539 $\times 10^{-2}$ \\ 
  & 7 & 0.3566 $\times 10^{-1}$  & 0.2551 $\times 10^{-1}$ & 0.3566 $\times 10^{-1}$ & 0.2550 $\times 10^{-1}$ \\ 
  & 8 & 0.4302                   & 0.3624                 & 0.4307                 & 0.3628                  \\ 
\end{tabular}
\end{ruledtabular}
\end{center}
\end{table*}



\end{document}